\documentstyle{mn}
\input{psfig}
\newcommand{\etal}{{et al.~}}

\newcommand{\lta}{\la}
\newcommand{\gta}{\ga}

\newcommand{\kmsmpc}{\>{\rm km}\,{\rm s}^{-1}\,{\rm Mpc}^{-1}}
\newcommand{\kms}{\>{\rm km}\,{\rm s}^{-1}}

\newcommand{\Msun}{\>{\rm M_{\odot}}}
\newcommand{\Lsun}{\>{\rm L_{\odot}}}

\newcommand{\mtol}{\>{\rm (M/L)_{\odot}}}

\newcommand{\apj}{ApJ}

\newcommand{\aj}{AJ}
\newcommand{\mnras}{MNRAS}
\newcommand{\aap}{A\&A}
\newcommand{\aaps}{A\&AS}

\newcommand{\nat}{Nature}

\newdimen\hssize
\newdimen\hdsize
\begin{document}

%%%%%%%%%%%%%%%%%%%%%%%%%%%%%%%%%%%%%%%%%%%%%%%%%%%%%%%%%%%%%%%%%%%%%%%%%%%%%%

\title{Dwarf Galaxy Rotation Curves and the Core Problem of Dark Matter Haloes}
\author[van den Bosch \& Swaters]
       {Frank C. van den Bosch$^{1,2}$\thanks{Hubble Fellow} and
        Rob A. Swaters$^3$\\
        $^1$Department of Astronomy, University of Washington, Seattle, 
         WA 98195, USA\\
        $^2$Max-Planck Institut f\"ur Astrophysik, Karl Schwarzschild
         Str. 1, Postfach 1317, 85741 Garching, Germany\\
        $^3$Carnegie Institution of Washington, Washington DC 20015, USA}

%%%%%%%%%%%%%%%%%%%%%%%%%%%%%%%%%%%%%%%%%%%%%%%%%%%%%%%%%%%%%%%%%%%%%%%%%%

\date{}

\pagerange{\pageref{firstpage}--\pageref{lastpage}}
\pubyear{2000}

\maketitle

\label{firstpage}

%%%%%%%%%%%%%%%%%%%%%%%%%%%%%%%%%%%%%%%%%%%%%%%%%%%%%%%%%%%%%%%%%%%%%%%%%%%%%

\begin{abstract}
  The   standard cold  dark   matter  (CDM) model   has  recently been
  challenged by the claim that dwarf galaxies  have dark matter haloes
  with constant density cores, whereas     CDM predicts haloes    with
  steeply  cusped     density  distributions.  Consequently,  numerous
  alternative dark matter candidates have  recently been proposed.  In
  this  paper, we   scrutinize   the observational evidence   for  the
  incongruity between dwarf galaxies and the CDM  model.  To this end,
  we  analyze  the  rotation curves of    20 late-type dwarf  galaxies
  studied by Swaters (1999).  Taking  the effects of beam-smearing and
  adiabatic contraction   into account, we  fit  mass models  to these
  rotation curves with dark matter haloes  with different cusp slopes,
  ranging from constant density  cores to $r^{-2}$ cusps.  Even though
  the  effects  of  beam  smearing   are  small  for these  data,  the
  uncertainties in  the  stellar mass-to-light  ratio and  the limited
  spatial sampling of the halo's density  distribution hamper a unique
  mass decomposition.  Consequently, the rotation curves in our sample
  cannot be  used  to discriminate between  dark  haloes with constant
  density cores  and $r^{-1}$ cusps.   We show that the dwarf galaxies
  analyzed here  are  consistent with  cold  dark  matter haloes in  a
  $\Lambda$CDM cosmology, and  that there is thus  no need  to abandon
  the idea  that dark matter  is cold and collisionless.  However, the
  data is also consistent with  any alternative dark matter model that
  produces  dark   matter haloes with   central  cusps less steep than
  $r^{-1.5}$.  In fact,  we argue that  based on existing  HI rotation
  curves alone at   best weak limits  can be  obtained on cosmological
  parameters and/or the  nature of the dark  matter. In order to  make
  progress,  rotation curves   with   higher spatial   resolution  and
  independent measurements of the  mass-to-light ratio of the disk are
  required.
\end{abstract}

%%%%%%%%%%%%%%%%%%%%%%%%%%%%%%%%%%%%%%%%%%%%%%%%%%%%%%%%%%%%%%%%%%%%%%%%%%


%%%%%%%%%%%%%%%%%%%%%%%%%%%%%%%%%%%%%%%%%%%%%%%%%%%%%%%%%%%%%%%%%%%%%%%%%%

\section{Introduction}
\label{sec:intro}

The  standard cosmological model  for  structure formation combines an
inflationary Universe with  hierarchical   growth of  structures  that
originate from small fluctuations in the cosmic mass distribution.  In
addition to baryonic   matter, this model requires  non-baryonic  dark
matter   and  possibly some form   of   vacuum energy or quintessence.
Unfortunately, the  nature of the  dark matter,  which is the dominant
mass component,  still remains unknown.  A  large number of candidates
have been proposed of  which cold dark matter (CDM)  has been the most
popular.   Because CDM particles   have a negligible thermal  velocity
with  respect to the Hubble flow,  the original phase-space density of
cold dark matter is extremely  high.  Numerical simulations have shown
that  a   small  fraction  of  this  material  remembers  its  initial
phase-space  density even after it collapses  to  form a bound object.
This low-entropy  material settles in  the centers  of virialized dark
haloes,  thus creating steeply cusped density  profiles, and causing a
large fraction  of  haloes to  survive as substructure   inside larger
haloes (e.g., Navarro, Frenk \&  White 1996; Fukushige \& Makino 1997;
Moore \etal 1998, 1999b; Ghigna \etal  1998; Klypin \etal 1999a; White
\& Springel 1999).

These characteristics of CDM haloes,  however, seem to disagree with a
number   of observations.  First,  the   number of sub-haloes around a
typical Milky Way  galaxy, as identified by  satellite galaxies, is an
order of magnitude smaller than predicted by  CDM (Kauffmann, White \&
Guiderdoni 1993;  Klypin \etal 1999b;   Moore \etal 1999a).  Secondly,
the observed rotation curves of dwarf and low surface brightness (LSB)
galaxies seem to indicate that their dark  matter haloes have constant
density  cores instead of steep  cusps (Flores  \& Primack 1994; Moore
1994; Burkert  1995; Burkert \& Silk  1997;  McGaugh \&  de Blok 1998;
Stil  1999;  Moore \etal  1999b; Dalcanton \&  Bernstein 2000; Firmani
\etal 2001).  In view of these discrepancies, numerous alternatives to
the CDM  paradigm have recently been   proposed.  These include broken
scale-invariance (hereafter BSI; Kamionkowski \& Liddle 2000; White \&
Croft 2000), warm dark  matter (hereafter WDM; Sommer-Larsen \& Dolgov
1999; Hogan \& Dalcanton   2000), scalar field dark  matter (hereafter
SFDM;  Peebles  \& Vilenkin 1999; Hu   \& Peebles 1999;  Peebles 2000;
Matos,  Siddharta  \&   Urena-L\'opez  2000),  and   various  sorts of
self-interacting or annihilating dark matter (hereafter SIDM; Carlson,
Machacek  \& Hall  1992;  Spergel  \&  Steinhardt 2000;  Mohapatra  \&
Teplitz 2000; Firmani \etal   2000; Goodman 2000; Kaplinghat, Knox  \&
Turner   2000; Bento \etal 2000).  Whereas   particle physics does not
prefer CDM over for  instance WDM, SFDM, or  SIDM, the former has  the
advantage  over   the    latter  that it    has   no free  parameters.
Furthermore,  most of  these alternatives  seem  unable to solve  both
problems simultaneously (Moore  \etal 1999b; Hogan \& Dalcanton  2000;
Col\'{i}n, Avila-Reese \& Valenzuela  2000; Dalcanton \& Hogan  2000),
and   often the   alternatives face   their own   problems (Spergel \&
Steinhardt 2000;   Hannestad  1999; Burkert  2000;   Moore \etal 2000;
Yoshida   \etal 2000; Kochanek   \& White  2000; Miralde-Escude  2000;
Sellwood 2000).

As an  alternative to  modifying the  nature  of the dark matter,  the
sub-structure and core   problems might   be solved once    additional
baryonic    physics are taken   into   account.  Several studies  have
suggested  that processes such  as reionization and supernova feedback
can help to suppress star  formation and to decrease central densities
in  low-mass dark  matter haloes  (e.g.,   Navarro, Eke \& Frenk  1996;
Gelato \&   Sommer-Larsen 1999;  van  den Bosch   \etal 2000; Bullock,
Kravtsov \& Weinberg 2000;   Binney, Gerhard \&  Silk 2001).   Whereas
these processes  may  indeed   help to   solve the  problem   with the
over-abundance of  satellite  galaxies, the suggestion  that  feedback
processes can actually  destroy  steep  central cusps seems   somewhat
contrived  in light of  more   detailed simulations. For instance,  as
shown by  Navarro, Eke \& Frenk, the  effects are only  substantial if
large fractions of  baryonic mass are   expelled, which seems  hard to
reconcile  with the low  ejection  efficiencies found in more detailed
hydro-dynamical simulations (e.g., Mac-Low \& Ferrara 1999; Strickland
\& Stevens 2000).

It is evident from the above discussion that the long-time popular CDM
paradigm is currently facing its biggest  challenge to date.  However,
before  abandoning CDM on the  grounds  that  it is inconsistent  with
observations,    it  is  worthwhile   to    more closely examine   the
observational evidence against it.   In this paper we scrutinize CDM's
most persistent   problem: the claim  that  the dark  haloes  of dwarf
galaxies,  as inferred  from their  rotation curves, are  inconsistent
with  CDM predictions.   The main   motivation  for this work is  that
recent work on the rotation curves of LSB galaxies has shown that once
data  with sufficient resolution  is obtained, or  the effects of beam
smearing  are properly taken into   account, the inner rotation curves
are  significantly steeper and allow   for more centrally concentrated
dark matter  haloes (Swaters 1999; van  den Bosch \etal 2000; Swaters,
Madore  \& Trewhella 2000).  In  fact, these studies  have pointed out
that,  in contrast with previous claims,  current data on LSB rotation
curves are consistent with CDM predictions.

Here we analyze    a set of  HI rotation   curves  of a sample of   20
late-type  dwarf  galaxies.     Taking beam  smearing  and   adiabatic
contraction of  the dark matter into   account, we investigate whether
the rotation curves of the galaxies in our  sample are consistent with
CDM haloes. Although we cannot rule  out that these dwarf galaxies have
dark haloes  with constant density  cores, we find  that their rotation
curves are  consistent with cold  dark matter haloes   as expected in a
$\Lambda$CDM cosmology.

\section{The data}
\label{sec:data}

The HI rotation curves  that we use in this   paper have been  derived
from the data  presented  in Swaters  (1999, hereafter  S99).  The  HI
observations  were done with the  Westerbork Synthesis Radio Telescope
(WSRT) as  part of the  Westerbork HI Survey  of Spiral  and Irregular
Galaxies  Project (WHISP, see Kamphuis,  Sijbring \& van Albada 1996).
The observations and  data reduction are discussed  in detail in  S99.
From the sample of 73 late-type  dwarf galaxies we only selected those
galaxies that according to  S99 have high  quality rotation curves (no
strong asymmetries  and   high  signal-to-noise ratio).  This   sample
consists  of 20 galaxies, which have  inclination  angles in the range
$39^\circ\leq i \leq 80^\circ$. Galaxies with $i > 80^\circ$ have been
excluded   from the sample because  they  require a somewhat different
analysis.

Table~\ref{tab:data} lists  the   properties of the  galaxies   in our
sample.   The absolute  magnitudes,  disk  scale lengths, and  central
surface brightnesses have   been determined  from $R$-band  photometry
(Swaters \& Balcells 2001,    hereafter SB01).  The distances are   as
adopted by SB01: where possible  stellar distance indicators have been
used, mostly brightest stars.  If these  were not available, distances
based on group membership were used,  or, if these were unavailable as
well,  the  distance was calculated  from   the HI systematic velocity
following the prescription   in Kraan-Korteweg (1986) with an  adopted
Hubble  constant of $H_0 = 75  \kmsmpc$.   $B-R$ colors, available for
only 6 galaxies, are also taken from SB01.
\begin{table*}
\begin{minipage}{10.5truecm}
\caption{Properties of sample of late-type dwarf galaxies.}
\label{tab:data}
\begin{tabular}{rrccccrccc}
  UGC &  $D$   &  $M_R$   & $\mu_0^R$ & $R_d$ & $V_{\rm last}$ &
 $N_V$ & $B-R$ & $i$ & $\Lambda$CDM \\
  731 &  $8.0$ & $-16.63$ & $23.0$ & $1.65$ & $74$ & $12$ &$0.85$ & $57$ & + \\
 3371 & $12.8$ & $-17.74$ & $23.3$ & $3.09$ & $86$ & $11$ &$1.08$ & $49$ & + \\
 4325 & $10.1$ & $-18.10$ & $21.6$ & $1.63$ & $92$ &  $8$ &$0.85$ & $41$ & + \\
 4499 & $13.0$ & $-17.78$ & $21.5$ & $1.49$ & $74$ &  $9$ & $--$  & $50$ & + \\
 5414 & $10.0$ & $-17.55$ & $21.8$ & $1.49$ & $61$ &  $7$ & $--$  & $55$ & ? \\
 6446 & $12.0$ & $-18.35$ & $21.4$ & $1.87$ & $80$ & $11$ & $--$  & $52$ & + \\
 7232 &  $3.5$ & $-15.31$ & $20.2$ & $0.33$ & $44$ &  $5$ &$0.81$ & $59$ & ? \\
 7323 &  $8.1$ & $-18.90$ & $21.2$ & $2.20$ & $86$ & $10$ & $--$  & $47$ & ? \\
 7399 &  $8.4$ & $-17.12$ & $20.7$ & $0.79$ &$109$ & $12$ &$0.78$ & $55$ & + \\
 7524 &  $3.5$ & $-18.14$ & $22.2$ & $2.58$ & $79$ & $31$ & $--$  & $46$ & + \\
 7559 &  $3.2$ & $-13.66$ & $23.8$ & $0.67$ & $33$ & $10$ & $--$  & $61$ & ? \\
 7577 &  $3.5$ & $-15.62$ & $22.5$ & $0.84$ & $18$ & $10$ & $--$  & $63$ & ? \\
 7603 &  $6.8$ & $-16.88$ & $20.8$ & $0.90$ & $64$ & $12$ & $--$  & $78$ & ? \\
 8490 &  $4.9$ & $-17.28$ & $20.5$ & $0.66$ & $78$ & $30$ & $--$  & $50$ & + \\
 9211 & $12.6$ & $-16.21$ & $22.6$ & $1.32$ & $65$ & $10$ & $--$  & $44$ & + \\
11707 & $15.9$ & $-18.60$ & $23.1$ & $4.30$ &$100$ & $13$ & $--$  & $68$ & + \\
11861 & $25.1$ & $-20.79$ & $21.4$ & $6.06$ &$153$ & $10$ & $--$  & $50$ & + \\
12060 & $15.7$ & $-17.95$ & $21.6$ & $1.76$ & $74$ &  $9$ & $--$  & $40$ & + \\
12632 &  $6.9$ & $-17.14$ & $23.5$ & $2.57$ & $76$ & $17$ &$0.91$ & $46$ & + \\
12732 & $13.2$ & $-18.01$ & $22.4$ & $2.21$ & $98$ & $15$ & $--$  & $39$ & + \\
\end{tabular}

\medskip

Column~(1) lists the  UGC  number of  the galaxy.  Columns~(2)  -- (6)
list the distance to the galaxy (in Mpc), absolute $R$-band magnitude,
central $R$-band  surface  brightness (in   mag arcsec$^{-2}$),  scale
length  of  the stellar  disk  (in kpc),   and  the  observed rotation
velocity  $V_{\rm   last}$ (in  $\kms$) at  the   last measured point.
Column~(7) lists the number of data  points, $N_V$, along the rotation
curve (two  data points per beam).  Columns~(8) and~(9) list the $B-R$
color  (if available) and the adopted  inclination angle (in degrees),
respectively.  Magnitudes and  central surface brightnesses have  been
corrected  for inclination   and   galactic extinction,  but not   for
internal   extinction.   Finally,  column~(10)  indicates whether  the
galaxy is consistent  with a $\Lambda$CDM  cosmology (+) or whether no
meaningful   fit  can  be    found  (?).    See   the  discussion   in
\S~\ref{sec:disc} for details.

\end{minipage}
\end{table*}

The original HI observations have been obtained with a typical beam of
$14''    \times  14''/\sin \delta$   (with     $\delta$ the   object's
declination).     In  general,  the    signal-to-noise  ratio at  this
resolution was too low to obtain reliable rotation curves.  Therefore,
the data were convolved to a  resolution of approximately $30'' \times
30''$.  Velocity fields were constructed by fitting Gaussian curves to
the   observed  line profiles at  each   position.  Next, the rotation
velocities    and their formal errors  were   determined  by fitting a
tilted-ring model to the velocity fields assuming constant inclination
and position   angles. Where possible     the orientation angles  were
determined from  the velocity fields,  and in the remaining cases from
the optical images. For details  about the determination of the tilted
ring parameters see S99. Note, however, that our analysis is different
from  the one presented  in S99,  where  an iterative method, based on
modelling of  the  observed  data  cubes, was used   to  approximately
correct the  rotation curve  for the effects  of beam  smearing.   The
rotation curves we use here, however, have not been corrected for beam
smearing.  Instead, we beam-smear   our models before comparison  with
the data, following the procedure detailed in \S~\ref{sec:method}.

\section{Rotation curve fitting}
\label{sec:fitrc}

\subsection{Mass components}
\label{sec:components}

For the mass  modelling presented in  this paper, we assume that there
are three main mass components in each galaxy: an infinitesimally thin
gas disk, a thick stellar disk, and a spherical dark halo.  We closely
follow the procedure outlined in  van den Bosch \etal (2000; hereafter
BRDB), which we briefly outline below for completeness.

In  order to  determine the contribution  of the  gas to  the galaxy's
circular velocity, we make the  assumption that the gas is distributed
axisymmetrically  in   an   infinitesimally  thin disk.   Under   this
assumption the circular velocities due to the  self-gravity of the gas
can be computed from the gas surface density using equation~[2-146] of
Binney  \&  Tremaine (1987). We  model  the HI density distribution as
follows:
\begin{equation}
\label{sbHI}
\Sigma_{\rm HI}(R) = \Sigma_0 \left({R \over R_1}\right)^{\beta}
{\rm e}^{-R/R_1} + f \; \Sigma_0 \, {\rm e}^{-((R-R_2)/\sigma)^2}.
\end{equation}
The first term is  identical  to the surface  density profile  used in
BRDB,  and represents an exponential disk  with scale length $R_1$ and
with  a central  hole,  the extent of  which depends  on $\beta$.  The
second term corresponds  to a Gaussian  ring  with radius $R_2$ and  a
FWHM $\propto \sigma$.  The flux ratio between these two components is
set by  $f$.   The form  of  equation~(\ref{sbHI})  has  no particular
physical motivation,  but  should be regarded as  a  fitting function. 
When computing the circular velocities of  the atomic gas, we multiply
$\Sigma_{\rm  HI}$ by a factor 1.3  to correct for the contribution of
helium.
 
For the stellar disk we  assume a thick exponential 
\begin{equation}
\label{rhostar}   
\rho^{*}(R,z) = \rho^{*}_0  \,  \exp(-R/R_d) \, {\rm sech}^2(z/z_0) 
\end{equation} 
where $R_d$ is the scale length of the disk.  Throughout we set $z_0 =
R_d/6$.  The   exact    value  of this   ratio,    however,  does  not
significantly  influence the  results.  The circular  velocity of  the
stellar disk  is computed using  equation~[A.17] in  Casertano (1983),
and   properly scaled with  the   stellar $R$-band mass-to-light ratio
$\Upsilon_R$.  None of  the galaxies in  our sample  has a significant
bulge component.
 
We assume that initially  the  dark and  baryonic matter virialize  to
form a spherical halo with a density distribution given by
\begin{equation}
\label{haloprof}
\rho(r) = {\rho_0 \over (r/r_s)^{\alpha} (1 + r/r_s)^{3-\alpha}},
\end{equation}
with $r_s$ being the scale radius of the halo, such that $\rho \propto
r^{-\alpha}$   for $r \ll r_s$ and   $\rho \propto r^{-3}$  for $r \gg
r_s$. For $\alpha = 1$   equation~(\ref{haloprof}) reduces to the  NFW
profile (Navarro, Frenk \& White  1997).  We define the  concentration
parameter $c = r_{200}/r_s$, with $r_{200}$ the radius inside of which
the mean density is 200 times the critical density for closure, i.e.,
\begin{equation}
\label{rvir}
{r_{200} \over h^{-1} {\rm kpc}} = {V_{200} \over \kms}.
\end{equation}
Here $V_{200}$ is the circular  velocity at $r_{200}$,  and $h = H_0 /
100  \kmsmpc$.  

The formation of  the stellar and gaseous disks  due to the cooling of
the baryons inside  the virialized halo leads to  a contraction of the
dark  matter component.   We make   the assumption  that  the baryonic
collapse is slow, and take this adiabatic contraction of the dark halo
into account following  the   procedure in  Barnes \&  White   (1984),
Blumenthal \etal (1986) and Flores \etal  (1993). The halo mass inside
radius $r$, required  for the  adiabatic contraction computations,  is
given by
\begin{equation}
\label{mhalo}
M_{\rm halo}(r) = M_{200} {\mu(xc) \over \mu(c)} 
\end{equation}
with $x = r/r_{200}$ and 
\begin{equation}
\label{fhalo}
\mu(x) = \int\limits_{0}^{x} y^{2 - \alpha} (1 + y)^{\alpha - 3} {\rm d}y
\end{equation}

\subsection{Beam smearing}
\label{sec:method}

As  is evident from  the  results presented in   S99  and BRDB, it  is
important that the effects   of beam-smearing are properly taken  into
account.  Rather than attempting to deconvolve the observations (which
is an ill-defined problem), we convolve our  models with the effective
point spread function $P$ (i.e., the beam) of the interferometer.  The
convolved surface brightness at a position $(x,y)$ on the plane of the
sky is
\begin{equation}
\label{convsig}
\tilde{\Sigma}(x,y) = \int\limits_{0}^{\infty} {\rm d}r \, r \, 
  \int\limits_{0}^{2 \pi} {\rm d}\theta \; \Sigma(r') \, 
  P(r,\theta - \theta_0).
\end{equation}
Here $r' = \sqrt{x'^2 + y'^2}$, where $x' = x + r \cos \theta$ and $y'
= (y + r \sin \theta) /  \cos i$ are  the Cartesian coordinates in the
equatorial plane of the disk, $i$ is the disk's inclination angle, and
$\theta_0$ is the angle  between the major axes of  the galaxy and the
beam (for  which we adopt  a two-dimensional  Gaussian, see BRDB). The
underlying    surface  brightness,    $\Sigma(R)$,    is   modeled  by
equation~(\ref{sbHI}). Note that we assume that the HI distribution is
axisymmetric.

Beam    smearing  also  affects    the observed    rotation velocities
$\tilde{V}_{\rm rot}$ at a position $(x,y)$ on the plane of the sky by
causing gas  from a larger  area  of the  disk  to contribute  to  the
observed line of sight velocity:
\begin{equation}
\label{convvel}
\tilde{V}_{\rm rot}(x,y)={1 \over \tilde{\Sigma}} 
  \int\limits_{0}^{\infty} {\rm d}r \, r \!\! \int\limits_{0}^{2 \pi}
  {\rm d}\theta \, \Sigma(r') V_{\rm los}(x',y') P(r,\theta - \theta_0) 
\end{equation} 
where   $V_{\rm los}$ is the line   of  sight velocity.  Throughout we
assume that the gas moves on circular orbits in  the plane of the disk
and has zero  intrinsic   velocity dispersion.  As argued  by  Swaters
(1999),  the asymmetric drift  corrections  as calculated from  the HI
distribution are generally small and have little effect on the derived
rotation curves.
 
\subsection{Fitting procedure}
\label{sec:fitting}

The first step in  fitting mass  models to  the rotation curves  is to
determine the best-fit model for the  true underlying HI distribution. 
The surface density distribution of equation~(\ref{sbHI}) has six free
parameters. Note, however, that $\Sigma_0$ is completely determined by
normalizing the models to the total mass in HI and can thus be ignored
in the fitting  routine.   We  determine the best-fit   parameters  by
minimizing
\begin{equation}
\label{chisb}
\chi^2_{\rm HI} = \sum_{i=1}^{N_{\rm HI}}\left( {\Sigma_{\rm obs}(R_i)
    - \tilde{\Sigma}(R_i) \over \Delta \Sigma_{\rm
      obs}(R_i)}\right)^2,
\end{equation}
with $\Sigma_{\rm obs}$ the observed HI density distribution at $30''$
resolution and   $\Delta \Sigma_{\rm obs}$ the   corresponding errors.
The     results  are  shown      in   the  upper-right     panels   of
Figures~\ref{fig:slopea}-\ref{fig:slopes}.  Open circles correspond to
the observed HI surface density and solid lines to the best-fit model.
In  most cases, the unsmeared HI  distribution of the best-fit models,
indicated by dashed lines, is fairly similar to that after convolution
with the beam, indicating that the  effects of beam-smearing for these
data are only small. In two cases, UGC~7524  and UGC~7603, our fitting
function (equation~[\ref{sbHI}])  can  not satisfactorily describe the
data.   In these  cases we use  the  data of the  full resolution (see
\S~\ref{sec:data}) as a model for the true underlying HI distribution.

Once $\Sigma_{\rm HI}$ is known we can  compute the beam-smeared model
rotation curves.   For a given choice of  the Hubble constant the mass
models described  above have four  free parameters  to  fit the  data:
$\Upsilon_R$,    $\alpha$,  $c$,    and $V_{200}$  (or    equivalently
$r_{200}$).   For  a given  ($\alpha$,$\Upsilon_R$)  we  determine the
best-fitting $c$ and $V_{200}$ by minimizing
\begin{equation}
\label{chivel}
\chi^2_{\rm vel} = \sum_{i=1}^{N_{\rm vel}}\left( {V_{\rm obs}(R_i) -
    \tilde{V}(R_i) \over \Delta V_{\rm obs}(R_i)}\right)^2.
\end{equation}
Here $\Delta V_{\rm obs}$ are the formal errors  on $V_{\rm obs}$ from
the tilted-ring   model  fits, and $\tilde{V}(R_i)$  is  computed from
equation~(\ref{convvel}) with $x=R_i$ and $y=0$.

\subsection{Uncertainties on the rotation velocities}
\label{sec:errors}

An important  issue  in constraining  the  density distribution of the
dark matter  haloes is how  to interpret  $\chi^2_{\rm vel}$.  One  can
only    use  the absolute  values   of  $\chi^2_{\rm  vel}$ to compute
confidence levels  for our models, if  the errors $\Delta V_{\rm obs}$
are the proper, normally  distributed errors, there are  no systematic
errors, the  data points are  independent, the  assumptions underlying
the model are correct, and  the mass-model is a proper  representation
of the real mass  distribution.   However, the  fact that  the minimum
$\chi^2_{\rm vel}$ differs considerably  from the number of degrees of
freedom for almost  all galaxies indicates that  we do  not meet these
criteria. This does not come  as a surprise.   First of all, errors in
the assumed  inclination  angle,  distance, beam  parameters,  and the
distribution  of gas  and stars all  lead  to systematic errors in the
dark matter  density  distribution.   Furthermore, there are  numerous
assumptions underlying our mass-models, each of which may be in error.
For instance, we  assume that the halo  is sperical, that  the disk is
axisymmetric and that the gas moves on  perfectly circular orbits with
zero  intrinsic velocity dispersion   (i.e., we thus ignore asymmetric
drift).   In   addition,  we assume   that $\Sigma_{\rm   gas}   = 1.3
\Sigma_{\rm  HI}$  and that  $\Upsilon_R$  is constant  throughout the
stellar  disk.  We thus ignore any  contribution from molecular and/or
ionized   gas as well  as any  radial  changes  in stellar population.
Given all   these   potential  sources of  confusion,   we    only use
$\chi^2_{\rm vel}$ to  assess the {\it relative}  quality of the model
fits.  We do not  try to assign any confidence  levels to the absolute
values of either  $\chi^2_{\rm  vel}$  or $\Delta  \chi^2_{\rm  vel}$.
Henceforth, if, for instance, a model for a particular galaxy yields a
smaller $\chi^2_{\rm vel}$ for $\alpha =  0$ than for $\alpha =1$, the
model with a constant density core provides a  better fit than the NFW
model, but it  does not necessarily  imply that the rotation curve  is
{\it  inconsistent}  with CDM. For   these reasons,   we use  physical
criteria (discussed in \S~\ref{sec:disc})  rather than criteria  based
on uncertain confidence levels to assess whether a model is consistent
with the data.

\subsection{Degeneracies in the mass modelling}
\label{sec:degen}

Before interpreting the results in terms of constraints on the density
distribution of dark matter haloes it is useful to examine some models.
To that end, we construct three model  galaxies moulded after UGC~731,
i.e., the models have the same HI and stellar disks as UGC~731, and we
adopt   the   same       distance  and   inclination    angle     (see
Table~\ref{tab:data}).  We  add  a dark  matter  halo  with  a density
distribution   given   by equation~(\ref{haloprof})   and compute  the
beam-smeared   rotation    velocities   at  $15''$  intervals.    This
corresponds  to roughly half   the beam size and   is identical to the
interval between actual data points of the  rotation curve of UGC~731.
We  convolve  the velocity field  using the   same  beam-size and beam
orientation as for the  true UGC~731 data.   Finally, we add a  random
Gaussian error to the model rotation velocities (with variance $\Delta
V$).  All three models have the same mass  distribution: $\alpha = 1$,
$c=20$, $V_{200} =  75 \kms$, and $\Upsilon_R  = 2.0 \mtol$.   What we
vary, however, is the way the rotation curve is sampled: Model~1 has a
rotation  curve with $6$ data  points  extending out to 0.06 $r_{200}$
and  with $\Delta V  = 2.0 \kms$.  The  rotation  curve of model~2 has
twice  as many data points,  thus extending twice  as far out, and has
the same  $\Delta V$.   Model~3, finally,  has  a rotation  curve that
extends equally  far  as that of  model~2, but  with $\Delta  V  = 0.2
\kms$.  For  comparison, the real data of  UGC~731 consists of 12 data
points with $\langle \Delta V \rangle = 1.4 \kms$, and model~2 is thus
a  fair representation of  the actual data.   Also, as can  be seen in
column~(7) in Table~\ref{tab:data}, the number  of data points for the
other   galaxies in our  sample   varies from  $5$  (UGC~7232) to $31$
(UGC~7524). Note  that,  because of  distance effects,  a larger $N_V$
does not necessarily mean that the density distribution of the halo is
probed to larger radii.
\begin{figure*}
\psfig{figure=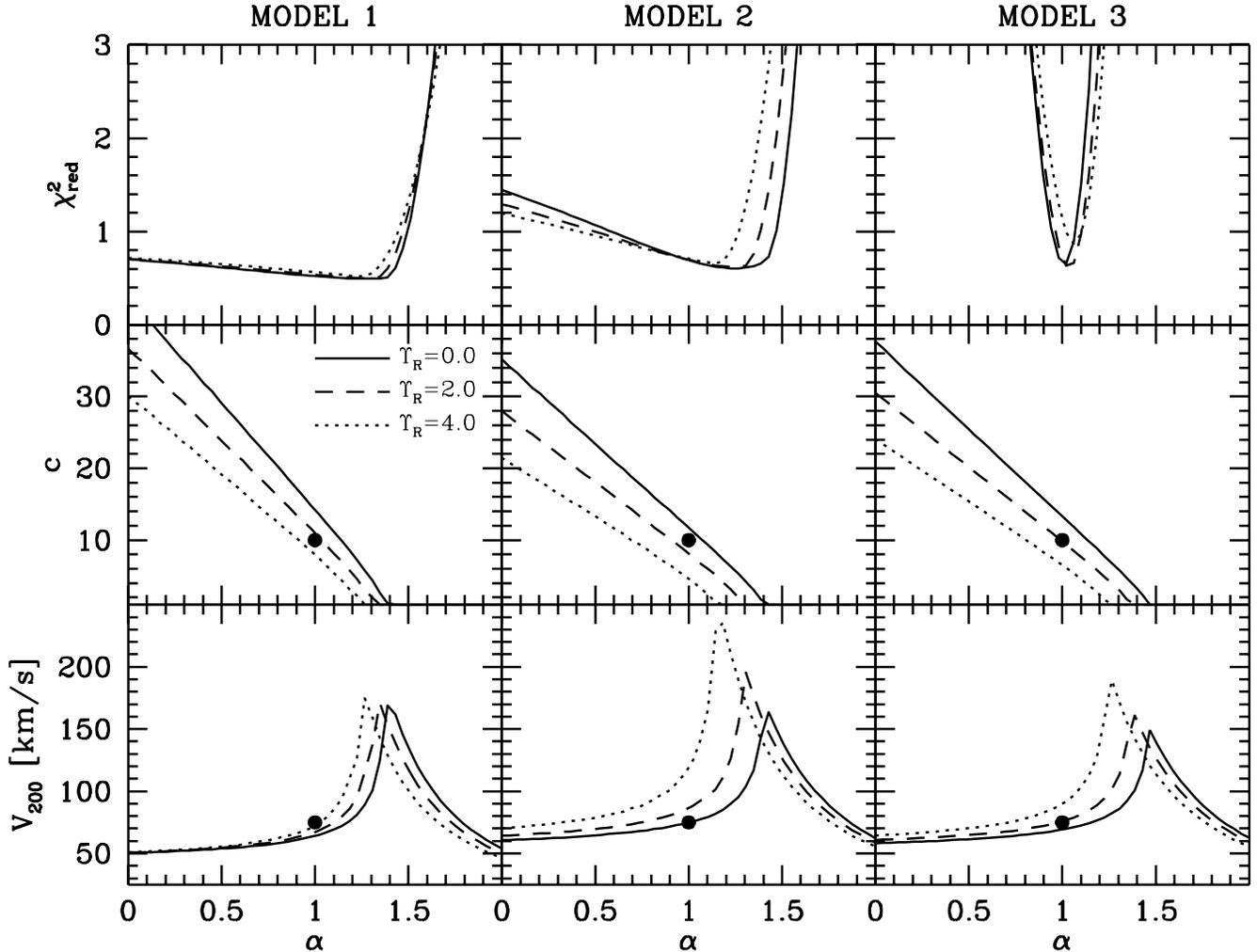,width=\hdsize}
\caption{Results of  analyzing  three model rotation
  curves,  moulded after  UGC~731.   All models  have the same density
  distribution (indicated by a  black dot); they   only differ in  the
  extent  of the   rotation curve  and the   errors  on the   rotation
  velocities, as indicated in the text.  In fitting the model rotation
  curves three different   mass-to-light  ratios  have   been assumed:
  $\Upsilon_R = 0$, $\Upsilon_R = 2.0 \mtol$ (which is the input value
  of the models),  and $\Upsilon_R = 4.0 \mtol$.   The labeling is  as
  indicated in   the left  middle  panel. Note  that  the  mass models
  derived   from the rotation   curves  of  models~1 and~2  (both with
  $\Delta V = 2 \kms$) are strongly degenerate.  The rotation curve of
  model~3, for which $\Delta  V = 0.2 \kms$,  however, allows a fairly
  accurate    recovery   of the   input    mass   model, although  the
  mass-to-light ratio degeneracy remains.}
\label{fig:models}
\end{figure*}

We analyze our model rotation curves in the same way as we analyze the
data  for    the   dwarf galaxies.   The     results are    shown   in
Figure~\ref{fig:models}.  A number  of general trends, which are  also
present       in         the           real         data          (see
Figures~\ref{fig:slopea}-\ref{fig:slopes}), are  immediately apparent.
First  of all,  the  halo   concentration  $c$ of the   best-fit model
decreases  with increasing $\Upsilon_R$ and  $\alpha$.  This is easily
understood in terms of the total enclosed mass which has to be similar
for different models.  Secondly,  there is an $\alpha_{\rm  crit}$ for
which  $V_{200}$ is  maximal and $c=1$.     For $\alpha >  \alpha_{\rm
crit}$ the best fitting halo concentration  $c < 1$, implying that the
scale radius  $r_s$  is  larger  than   the  virial radius  $r_{200}$.
Clearly, for    $c<1$,  equation~(\ref{haloprof})  no   longer   is an
appropriate  description  of   the dark matter   distribution, and  we
therefore demand that   $c   \geq 1$.  Consequently,  for    $\alpha >
\alpha_{\rm crit}$ the halo concentration $c=1$ and the quality of the
fits rapildy decreases (i.e., $\chi^2$ increases).

In addition  to  these trends,   a large amount   of non-uniqueness is
apparent,  which can   be associated with   two distinct degeneracies.
First of all,  for any given cusp  slope $\alpha$, the relative amount
of  mass in the  stellar disk can be  traded off against the amount of
mass in the halo, while maintaining virtually equally good fits to the
data, i.e., different combinations of ($\Upsilon_R, c, V_{200}$) yield
similar  values of $\chi^2_{\rm  vel}$.  This degeneracy is well-known
from  the rotation curves of  high  surface brightness galaxies (e.g.,
van   Albada  \etal  1985),   and is   generally referred   to  as the
mass-to-light ratio degeneracy.   The second degeneracy  is that for a
given  mass-to-light  ratio $\Upsilon_R$,  different   combinations of
($\alpha, c, V_{200}$) with $\alpha \lta  1.5$ yield virtually equally
good fits  to the data,  unless the  errors  on the observed  rotation
velocities are  sufficiently  small   (cf.    models~2  and~3).   This
degeneracy, which   we refer  to  as  the cusp-core  degeneracy,  is a
consequence of  the fact  that  rotation curves   only sample a  small
fraction of the  dark matter density  distribution  (see also Lake  \&
Feinswog 1989).

\begin{figure}
\psfig{figure=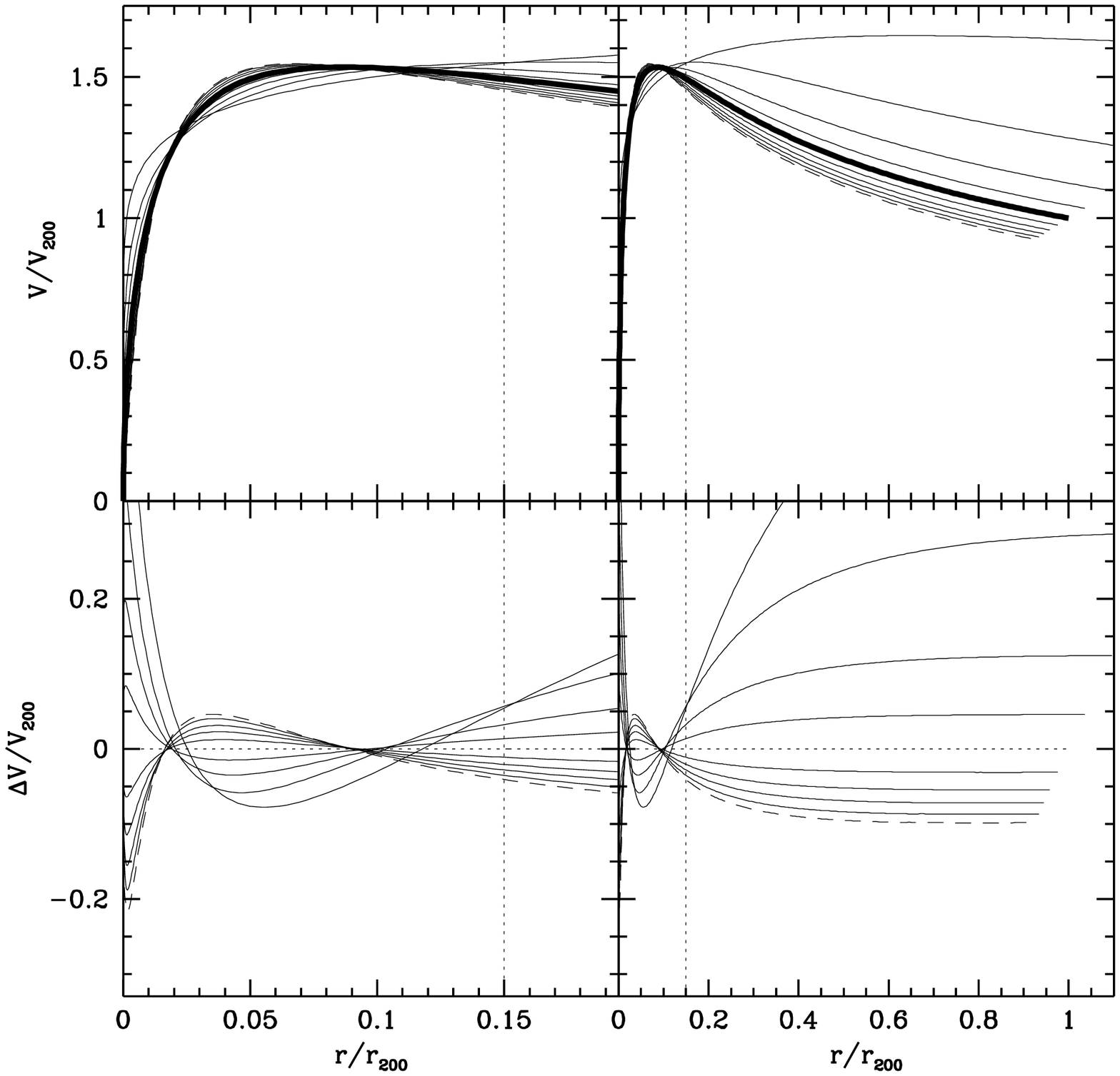,width=\hssize}
\caption{An illustration of the cusp-core  degeneracy. The thick solid
  lines in the upper two panels correspond to the circular velocity of
  a halo with a density distribution of equation~(\ref{haloprof}) with
  $\alpha = 1.0$  and   $c = 25$.   The   abscissa  and ordinate   are
  normalized to $r_{200}$ and $V_{200}$  of this density distribution,
  respectively, The  thin lines  are  best-fits to the  inner parts of
  this  circular  velocity curve of  models  with  $\alpha = 0.0, 0.2,
  0.4,...,1.8$, once again normalized to   $r_{200}$ and $V_{200}$  of
  the $\alpha = 1.0$ model. For clarity, the model with $\alpha = 0.0$
  is indicated by dashed lines.  The  lower panels plot the normalized
  differences, $(V_{\alpha} - V_{\alpha =  1})/V_{200}$ as function of
  $r/r_{200}$.  When fitting  the  models, only the velocities  out to
  $r_{\rm max}  = 0.15 r_{200}$ (indicated  by vertical  dotted lines)
  are taken into account.}
\label{fig:degen}
\end{figure}

To illustrate the nature of this cusp-core degeneracy we construct the
circular velocity curve, $V_{\alpha=1}(r)$, of a dark matter halo with
$\alpha=1$, $c=25$, and  $V_{200} = 100 \kms$.   Next, for  a range of
values for $\alpha$, we seek the values of $c$ and $V_{200}$ for which
$V_{\alpha}(r)$ (with $\alpha  \neq 1.0$) best  fits $V_{\alpha=1}(r)$
out to  a certain  radius $r_{\rm   max}$.  The results  are  shown in
Figure~\ref{fig:degen} for $r_{\rm max}  = 0.15 \, r_{200}$ (indicated
by a dotted  vertical  line).  The thick  curve  in the  upper  panels
corresponds to  $V_{\alpha=1}(r)$, normalized to  $V_{200}$.  The thin
curves  correspond to the   best-fitting $V_{\alpha}(r)$ for $\alpha =
0.0, 0.2,  0.4,...,1.8$.   As  is evident  from  the  left  panels  in
Figure~\ref{fig:degen}, where we plot the circular velocities only out
to $0.2 \, r_{200}$, the different  $V_{\alpha}(r)$ curves are in fact
very  similar.  Only for $\alpha  \gta 1.5$ does $V_{\alpha}(r)$ start
to deviate  more significantly  from $V_{\alpha=1}(r)$.  This explains
why  the  reduced $\chi^2_{\rm  vel}$    of models~1  and~2  increases
strongly for $\alpha \gta  1.5$.  For $\alpha  \lta 1.5$  the circular
velocity curves out to $r_{\rm max}$  are remarkably similar, and only
very accurate  rotation  curves  (i.e., with   small $\Delta V$)   can
discriminate between   the different  curves  (cf.  models~2   and~3).
Alternatively, accurate constraints on the actual density distribution
of  the dark  matter  requires a  rotation  curve that either  extends
sufficiently far,  or that has  sufficient independent measurements at
very  small  radii.  This  is  evident from  the  lower  two panels of
Figure~\ref{fig:degen},  which  plot    the     normalized  difference
$(V_{\alpha}   - V_{\alpha   = 1})  /  V_{200}$  as   function of  the
normalized radius,  and  which  show how   the  different $V_{\alpha}$
curves   diverge for both $r/r_{200}  \gta   0.2$ and $r/r_{200}  \lta
0.02$.  Unfortunately, in practice HI rotation curves rarely extend to
large enough radii,  do not have enough  spatial  resolution, have too
large errors, and suffer too much from systematic effects to lift this
cusp-core degeneracy.

\section{Results}
\label{sec:results}

The   results   for   each  individual     galaxy  are presented    in
Figures~\ref{fig:slopea}-\ref{fig:slopes} and discussed in some detail
in  the Appendix.   For UGC~7557   no converging  fit  to the observed
rotation curve  could  be achieved  with the   mass-model described in
Section~\ref{sec:fitrc}, and therefore no results are plotted for this
galaxy  (see discussion in the  Appendix).   The four  left panels  of
Figures~\ref{fig:slopea}-\ref{fig:slopes}  show   (from top to bottom)
$\chi^2_{\rm red}$,  $c$,  $V_{200}$,   and  the corresponding  baryon
fraction $f_{\rm bar}$ (see  \S~\ref{sec:degen}), all as a function of
the  cusp  slope  $\alpha$,    and   for three different   values   of
$\Upsilon_R$.   Here $\chi^2_{\rm red}  \equiv \chi^2_{\rm vel}/N_{\rm
df}$, with $N_{\rm  df}$ the number of degrees  of freedom.  The  same
general trends we found  for the model galaxies  are also  apparent in
the data, i.e.,  the halo concentration  $c$ decreases with increasing
$\Upsilon_R$ and $\alpha$,  and above a  certain value of $\alpha$ the
quality of the fits decreases rapidly while $c=1$.  As we indicated in
\S~\ref{sec:data} we   have used data  that is  convolved  to  a lower
resolution in order to   enhance the signal-to-noise ratio.    We have
checked  that our  best fit   models are  consistent   with the higher
resolution data and found good agreement.

As we have argued in \S~\ref{sec:errors} above,  we can not simply use
$\chi^2_{\rm red}$ to  put confidence  levels  on the  various models.
Furthermore, we have pointed out that two distinct degeneracies hamper
a unique mass-decomposition, which  is readily apparent from the  fact
that   models with very   different cusp  slopes and/or  mass-to-light
ratios yield roughly equally  good fits  (see for instance  UGC~11707,
UGC~12060, and UGC~12632).  However, some  constraints can be  imposed
by only    considering models  that   are  physically  realistic.  For
instance,  models with $\Upsilon_R   =  0$  are unrealistic, and   are
therefore not considered meaningful  model fits.  In addition,  we can
use the baryon fraction of each model to  check its physical validity.
For  each model we  compute  the baryonic  mass fraction  $f_{\rm bar}
\equiv (M_{\rm gas} +  M_{\rm stars})/M_{200}$ with $M_{200} = r_{200}
V^2_{200} /    G$ the total   mass of  the   galaxy (baryons plus dark
matter).  For  currently  popular cosmologies  with $\Omega_0=0.3$ and
$h=0.7$, and using recent  Big Bang nucleosynthesis constraints on the
baryon density   ($\Omega_{\rm bar} = 0.02   h^{-2}$; Burles \& Tytler
1998), one expects  a universal  baryons  fraction of roughly  $0.14$.
Note that  because  we ignore any  molecular  and/or ionized  gas, and
because feedback may  drive galactic winds and  expel baryons from the
halo,  $f_{\rm bar}$  may be  significantly  lower  than the universal
value.  In  order to leave  some room  for  the uncertainties  in  the
cosmological parameters, in what follows we shall only consider models
unrealistic if $f_{\rm bar} > 0.2$.  In section~5 below we address the
effects a  possible lower bound  on $f_{\rm  bar}$  might have  on our
conclusions.

\begin{figure*}
\psfig{figure=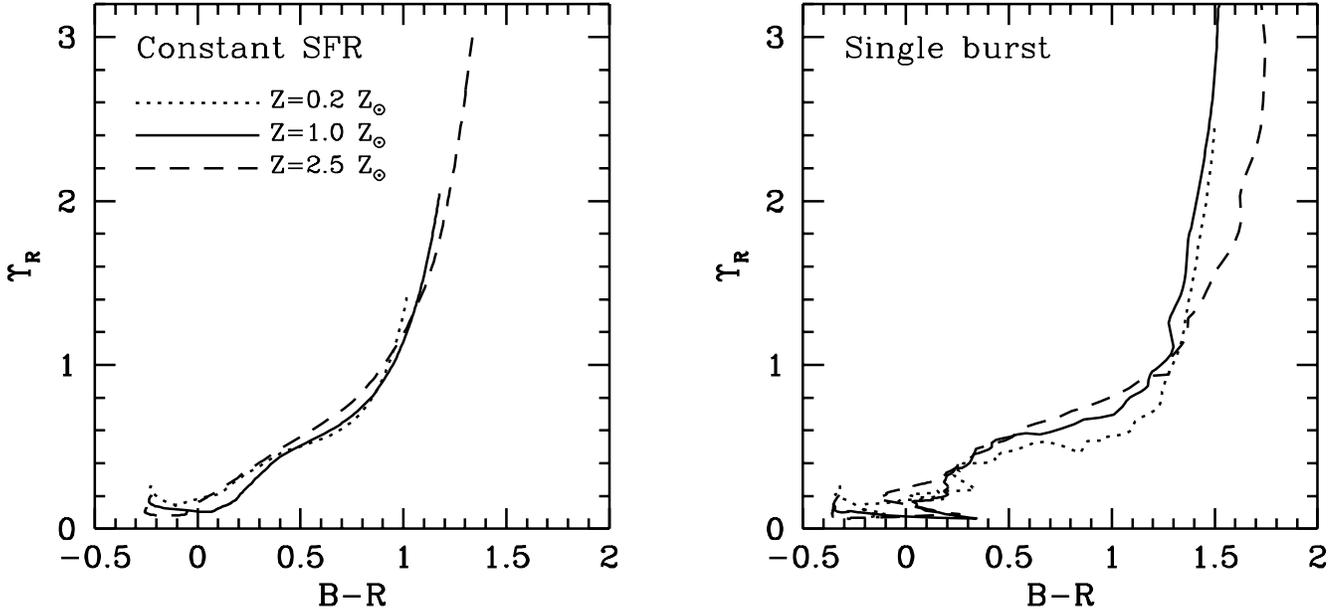,width=\hdsize}
\caption{The $R$-band mass-to-light ratio  $\Upsilon_R$ as function of
  the $B-R$ color   of Bruzual \&  Charlot  (1993)  stellar population
  models with  a Scalo IMF and  either a constant star  formation rate
  (left panel) or   a single burst  of star   formation (right panel).
  Results are shown for three different metallicities: one fifth Solar
  (dotted lines),  Solar (solid lines) and   two and half  times Solar
  (dashed  lines).  For  the dwarf  galaxies in our  sample with known
  $B-R$ we find $\langle  B-R \rangle =  0.87 \pm 0.09$, which implies
  that $\Upsilon_R \simeq 0.5 - 1.1 \mtol$.}
\label{fig:colups}
\end{figure*}

We can put some further constraints  on the mass models by considering
what  range  of $R$-band mass-to-light   ratios to expect.   Using the
Bruzual \& Charlot  (1993) stellar population  models we have computed
$B-R$ and  $\Upsilon_R$ for a Scalo  (1986) IMF and two different star
formation histories. Figure~\ref{fig:colups} plots $\Upsilon_R$ versus
$B-R$ for  three different metallicities  and for both a constant star
formation rate  (left panel)  and a  single burst  stellar  population
(right panel).  For the  six  galaxies in  our  sample for which  SB01
obtained $B$-band photometry we  find $\langle B-R  \rangle = 0.87 \pm
0.09$  (see  Table~\ref{tab:data}).  If  we  assume  that  there is no
internal    extinction in these   galaxies,   this  implies  $0.5 \lta
\Upsilon_R \lta  1.1$ for the  stellar  population models investigated
here.    This ignores the  contribution  of  any non-luminous baryonic
component that may have the same radial distribution as the stars, and
which     would   increase    the       mass-to-light   ratio.      In
Table~\ref{tab:results} we list the  parameters of the best-fit models
with $\alpha=1$ for both $\Upsilon_R=0$  and $\Upsilon_R = 1.0 \mtol$.
Although models with $\Upsilon_R =  0$ are unrealistic, these best-fit
models  yield a useful upper limit  on $c$ (cf.  Pickering \etal 1997;
Navarro 1998).  The models with $\Upsilon_R = 1.0 \mtol$ are chosen to
represent a typical mass-to-light  ratio. Furthermore, a comparison of
two models  with  different mass-to-light ratios   sheds  light on the
(non)-uniqueness of the mass models.

The    panels        in      the       lower-right   corners        of
Figures~\ref{fig:slopea}-\ref{fig:slopes},  plot the observed rotation
curves  (open circles with errorbars) together   with four models with
$\Upsilon_R = 1.0 \mtol$.  These are the best-fit models for $\alpha =
0$   (solid    lines),   $\alpha=0.5$  (dotted    lines), $\alpha=1.0$
(short-dash lines), and  $\alpha=1.5$ (long-dash  lines).  These plots
illustrate the typical quality of the model fits and the dependence on
the cusp slope $\alpha$.  In most  cases the individual curves for the
four  models   can hardly   been  discerned,  further emphasizing  the
cusp-core degeneracy discussed above.
\begin{table}
\caption{Best-fit parameters for models with $\alpha = 1$.}
\label{tab:results}
\begin{tabular}{rcrrcccc}
 UGC & $\Upsilon_R$ & $c$ & $V_{200}$ & $f_{\rm bar}$ & $c_{\rm min}$
& $\langle c \rangle$ & $c_{\rm max}$ \\ 
  731 & $0.0$ & $16.0$ &  $51.3$ & $2.3\times 10^{-2}$ & $4.7$ & $15.6$ & $31.3$ \\
      & $1.0$ & $13.5$ &  $52.3$ & $2.8\times 10^{-2}$ & $4.7$ & $15.6$ & $31.1$ \\
 3371 & $0.0$ &  $9.5$ &  $68.6$ & $1.6\times 10^{-2}$ & $4.4$ & $14.6$ & $29.2$ \\
      & $1.0$ &  $8.0$ &  $69.8$ & $2.1\times 10^{-2}$ & $4.4$ & $14.5$ & $29.1$ \\
 4325 & $0.0$ & $30.9$ &  $53.5$ & $2.1\times 10^{-2}$ & $4.6$ & $15.5$ & $30.9$ \\
      & $1.0$ & $25.7$ &  $52.7$ & $4.1\times 10^{-2}$ & $4.6$ & $15.5$ & $30.9$ \\
 4499 & $0.0$ &  $9.0$ &  $58.1$ & $2.7\times 10^{-2}$ & $4.6$ & $15.2$ & $30.4$ \\
      & $1.0$ &  $1.6$ & $131.5$ & $3.7\times 10^{-3}$ & $3.7$ & $12.3$ & $24.7$ \\
 5414 & $0.0$ & $<1.0$ & $253.6$ & $1.8\times 10^{-4}$ & $3.0$ & $10.1$ & $20.2$ \\
      & $1.0$ & $<1.0$ & $128.8$ & $2.3\times 10^{-3}$ & $3.7$ & $12.4$ & $24.9$ \\
 6446 & $0.0$ & $17.4$ &  $52.0$ & $4.1\times 10^{-2}$ & $4.7$ & $15.6$ & $31.1$ \\
      & $1.0$ &  $7.0$ &  $59.5$ & $4.8\times 10^{-2}$ & $4.5$ & $15.1$ & $30.1$ \\
 7232 & $0.0$ &  $4.2$ & $116.0$ & $2.3\times 10^{-4}$ & $3.8$ & $12.8$ & $25.5$ \\
      & $1.0$ & $<1.0$ &  $59.6$ & $3.8\times 10^{-3}$ & $4.5$ & $15.1$ & $30.1$ \\
 7323 & $0.0$ &  $4.7$ & $129.3$ & $1.4\times 10^{-3}$ & $3.7$ & $12.4$ & $24.9$ \\
      & $1.0$ & $<1.0$ & $193.9$ & $1.5\times 10^{-3}$ & $3.3$ & $11.0$ & $22.0$ \\
 7399 & $0.0$ & $23.1$ &  $62.8$ & $1.3\times 10^{-2}$ & $4.5$ & $14.9$ & $29.8$ \\
      & $1.0$ & $15.2$ &  $70.9$ & $1.4\times 10^{-2}$ & $4.4$ & $14.5$ & $29.0$ \\
 7524 & $0.0$ &  $8.5$ &  $71.9$ & $1.1\times 10^{-2}$ & $4.3$ & $14.4$ & $28.9$ \\
      & $1.0$ &  $4.8$ &  $82.5$ & $1.4\times 10^{-2}$ & $4.2$ & $13.9$ & $27.9$ \\
 7559 & $0.0$ &  $1.4$ & $135.4$ & $1.4\times 10^{-4}$ & $3.7$ & $12.2$ & $24.4$ \\
      & $1.0$ &  $1.2$ & $135.1$ & $1.6\times 10^{-4}$ & $3.7$ & $12.2$ & $24.4$ \\
 7603 & $0.0$ &  $5.5$ &  $87.9$ & $3.0\times 10^{-3}$ & $4.1$ & $13.7$ & $27.4$ \\
      & $1.0$ & $<1.0$ & $217.3$ & $3.9\times 10^{-4}$ & $3.2$ & $10.6$ & $21.2$ \\
 8490 & $0.0$ & $24.2$ &  $50.3$ & $3.3\times 10^{-2}$ & $4.7$ & $15.6$ & $31.3$ \\
      & $1.0$ & $13.5$ &  $57.2$ & $2.9\times 10^{-2}$ & $4.6$ & $15.2$ & $30.5$ \\
 9211 & $0.0$ & $18.3$ &  $43.6$ & $5.6\times 10^{-2}$ & $4.9$ & $16.2$ & $32.4$ \\
      & $1.0$ & $14.8$ &  $44.5$ & $6.1\times 10^{-2}$ & $4.8$ & $16.1$ & $32.2$ \\
11707 & $0.0$ & $13.1$ &  $67.3$ & $5.3\times 10^{-2}$ & $4.4$ & $14.7$ & $29.3$ \\
      & $1.0$ & $11.2$ &  $66.9$ & $7.1\times 10^{-2}$ & $4.4$ & $14.7$ & $29.3$ \\
11861 & $0.0$ & $16.0$ & $106.3$ & $2.6\times 10^{-2}$ & $3.9$ & $13.1$ & $26.1$ \\
      & $1.0$ & $12.5$ & $100.6$ & $7.6\times 10^{-2}$ & $4.0$ & $13.2$ & $26.4$ \\
12060 & $0.0$ & $34.6$ &  $46.6$ & $7.8\times 10^{-2}$ & $4.8$ & $16.0$ & $31.9$ \\
      & $1.0$ & $24.3$ &  $46.5$ & $1.1\times 10^{-1}$ & $4.8$ & $16.0$ & $31.9$ \\
12632 & $0.0$ & $14.0$ &  $51.6$ & $2.8\times 10^{-2}$ & $4.7$ & $15.6$ & $31.1$ \\
      & $1.0$ & $12.6$ &  $51.4$ & $3.7\times 10^{-2}$ & $4.7$ & $15.6$ & $31.1$ \\
12732 & $0.0$ & $10.3$ &  $67.6$ & $5.3\times 10^{-2}$ & $4.4$ & $14.7$ & $29.3$ \\
      & $1.0$ &  $6.8$ &  $73.3$ & $4.7\times 10^{-2}$ & $4.3$ & $14.3$ & $28.7$ \\
\end{tabular}

\medskip

For each galaxy  we list the parameters  of  two best-fit models  with
$\alpha = 1$ (i.e., a dark matter halo  with a $r^{-1}$ density cusp):
one with $\Upsilon_R = 0$ and the other with $\Upsilon_R = 1.0 \mtol$.
In addition to the best-fit parameters $c$, $V_{200}$ (in $\kms$), and
the resulting  baryon  fraction $f_{\rm bar}$,  we  list the mean halo
concentration,  $\langle c \rangle$, for   $\Lambda$CDM haloes with the
same $V_{200}$ as the best-fit model, as well as  the $2 \sigma$ lower
and upper limits ($c_{\rm  min}$  and $c_{\rm max}$,  respectively) of
the distribution of $c$.  These values are computed using the model of
Bullock   \etal   (1999)  for     the  $\Lambda$CDM   cosmology   used
here.  Consistency with $\Lambda$CDM requires  that $c_{\rm min} < c <
c_{\rm max}$ and $f_{\rm bar} < 0.2$.

\end{table}

\section{Implications for the nature of the dark matter}
\label{sec:disc}

The main goal of this paper is  to assess whether  or not the rotation
curves  of dwarf  galaxies  are  consistent  with CDM.  Before  we can
address this, we need to define what  ``consistent with CDM'' means in
the context of  the density distribution of  dark haloes.  Ideally, one
would like  to place  confidence levels on  whether or  not a rotation
curve  is consistent  with  certain  CDM  predictions. However, as  we
discussed  in  \S~\ref{sec:errors},  we   can not  use    our $\chi^2$
statistic to compute such  confidence  levels.  Therefore we  follow a
different approach.  Several studies that have claimed inconsistencies
between (dwarf) galaxy rotation  curves and CDM haloes, indicated  that
when   a   model with  $\alpha=1.0$   is   fitted  the  implied   halo
concentration  $c$ is   too  low to   be  consistent  with CDM  (i.e.,
Pickering \etal 1997; Navarro    1998). We follow this   approach  and
investigate whether   the distribution  of   $c$ for  our best-fitting
models   with $\alpha =   1.0$ is consistent   with  predictions for a
particular CDM model.

High resolution $N$-body  simulations have shown that haloes  virialize
to density distributions of the form of equation~(\ref{haloprof}) with
$\alpha    \sim   1$.  Different simulations,   however,   often yield
different    values  for   the   concentrations.    Furthermore,   the
distribution of halo  concentrations is fairly  broad, and its  median
depends on the mass of the halo, its redshift, and the cosmology (Cole
\& Lacey 1996; Navarro,  Frank \& White  1996, 1997; Avila-Reese \etal
1999; Bullock \etal 1999; Jing 2000;  Jing \& Suto 2000).  Henceforth,
there is (currently) no well-defined boundary for $c$ to be considered
``consistent with  CDM''.  Instead, one can  only ask whether the {\it
statistical  properties of a sample  of rotation curves are consistent
with  CDM for a  given cosmology   and according to    a given set  of
simulations}.

In   what follows we    focus  on the  currently popular  $\Lambda$CDM
cosmology with $\Omega_0 = 0.3$, $\Omega_\Lambda  = 0.7$, $h=0.7$, and
$\sigma_8 = 1.0$.  The particular simulations  to which we compare our
results are presented in Bullock \etal (1999; hereafter B99), who also
give a simple recipe for computing $c$ and  its scatter as function of
halo mass\footnote{This model  uses somewhat different definitions for
  the  halo mass and  concentration, which we convert  to  the $c$ and
  $M_{200}$   used in  our  analysis.}.  The    reason for using  this
particular  recipe  is that it is  tested  against a large statistical
sample of  several thousand haloes. The expected  value of $c$ for each
galaxy (as  determined by  $V_{200}$  of the best-fit  model), and the
$2\sigma$ deviations  from the median,  indicated by $c_{\rm min}$ and
$c_{\rm max}$, are listed in Table~\ref{tab:results}.

We now define a  rotation curve to  be consistent with $\Lambda$CDM if
for $\alpha=1.0$ and  $0.5   \leq \Upsilon_R \leq   1.1$  there  is  a
best-fit model with $c_{\rm min} \leq  c \leq c_{\rm max}$ and $f_{\rm
bar}  \leq 0.2$.  These limits on  $\Upsilon_R$ and  $f_{\rm bar}$ are
motivated  in \S~\ref{sec:results}.  Note  that we do  not demand that
the models   with $\alpha = 1.0$ yield   the best  fit  (i.e., minimum
$\chi^2_{\rm red}$) of all models.  Instead, we demand that there is a
model  with   $\alpha=1.0$ for which     the resulting parameters  are
realistic.   According to this definition, we  find that 14 out of the
20  dwarf galaxies  in our  sample  are consistent  with $\Lambda$CDM.
These  galaxies   are indicated     by   a  `+'  in  column~(10)    in
Table~\ref{tab:data}.
\begin{figure*}
\psfig{figure=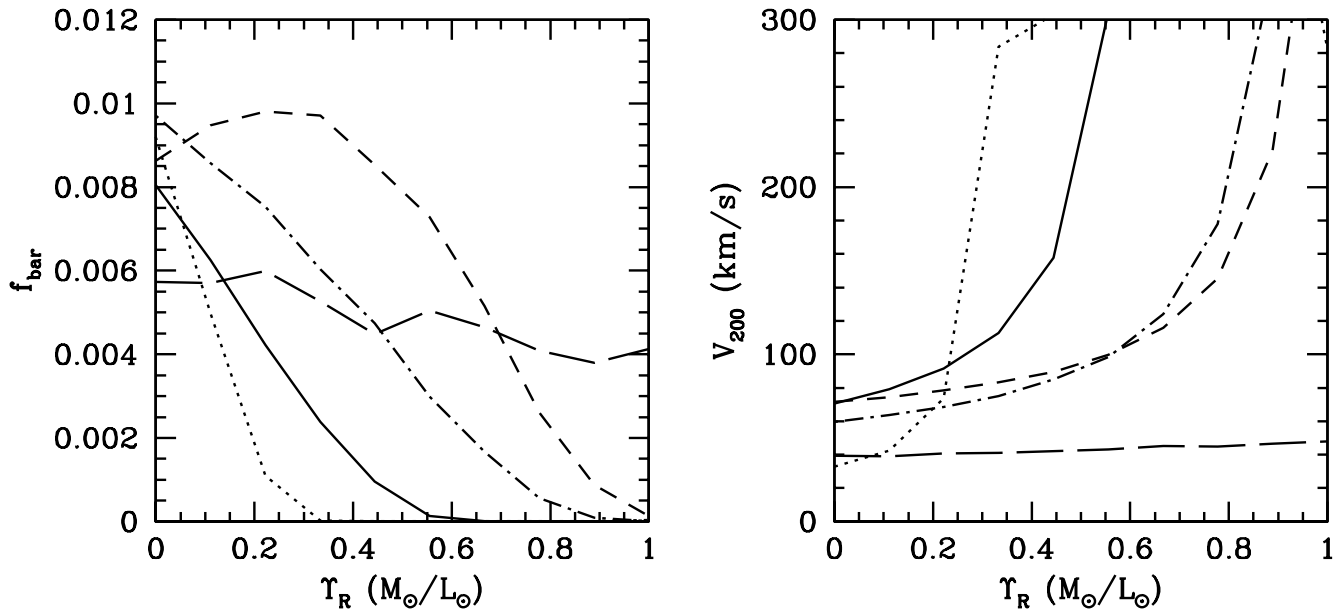,width=\hdsize}
\caption{The baryon fractions  (left) and virial velocities (right) as
  function of $\Upsilon_R$ for the five galaxies that are inconsistent
  with $\Lambda$CDM  haloes: UGC~5414  (solid lines), UGC~7232  (dotted
  lines),  UGC~7323 (short-dashed lines), UGC~7559 (long-dashed lines)
  and UGC~7603 (dot-dashed lines).  The  results plotted here are  for
  $\alpha=0$   (haloes with constant   density  cores) only.  Note  how
  $f_{\rm bar}$ deceases rapidly  with increasing  $\Upsilon_R$, while
  $V_{200}$  rapidly  increases.  For  $\Upsilon_R >  0.5 \Msun/\Lsun$
  (which  is  the more  realistic   regime), $f_{\rm bar} <   8 \times
  10^{-3}$, while $V_{200} > 90  \kms$ (except for UGC~7559).Given our
  current  understanding of feedback,  such  low baryon fractions  are
  hard to   reconcile with the  large  halo masses, and   we therefore
  consider these rotation curve  fits  unrealistic (see discussion  in
  text).}
\label{fig:nofit}
\end{figure*}

But what  about the other 6  galaxies?  For UGC~7577 no converging fit
could  be  obtained at all,  and  this  galaxy  is discarded from  the
following discussion.  The best fit models of  the other five galaxies
all  have  $\alpha=0.0$  (constant density  core)  and  $\Upsilon_R=0$
(which is   unphysical).  In   Figure~\ref{fig:nofit} we plot  $f_{\rm
bar}$   and   $V_{200}$  for  these  five   galaxies   as functions of
$\Upsilon_R$, whereby $\alpha$ is  kept constant at  zero.  As  can be
seen, the inferred   values  of $f_{\rm  bar}$ rapidly  decrease  with
increasing  $\Upsilon_R$, while  $V_{200}$  rapidly  increases.    For
$\Upsilon_R \gta 0.5 \Msun/\Lsun$, which is the more realistic regime,
one finds values of $f_{\rm bar}$ well below $0.01$.  With a universal
baryon fraction of $\sim 0.14$ this implies that $> 90$ percent of the
available baryons would have to  be expelled from  the disk.  Although
current   understanding of  feedback   is very  limited,   and  we are
reluctant to impose stringent lower limits on $f_{\rm bar}$, such high
ejection  efficiencies seem   inconsistent   with the high values   of
$V_{200}$ inferred  from the best-fit  models (see e.g., Dekel \& Silk
1986;  Efstathiou 2000).     In particular,  recent    hydro-dynamical
simulations have  indicated   that even  starburst  driven  winds  are
extremely inefficient in expelling  matter from systems with such high
virial  velocities (Mac Low  \&  Ferrara 1999;  Strickland  \& Stevens
2000).  Thus, unless a mechanism can be  devised that can expell $\gta
90$ percent of the available baryons from haloes with $V_{200} \gta 90
\kms$, we  conclude  that   for  these five   galaxies  none   of  the
($\alpha$,$\Upsilon_R$)-models  are realistic.    This  implies   that
whereas  the rotation curves  of  these galaxies are inconsistent with
the $\Lambda$CDM model, they do  not support an alternative picture in
which, for instance, dark  matter haloes have constant density  cores.
For these galaxies either (1) our  mass-model is inadequate, (2) there
are   systematic errors  in the  data,   or  (3) one  or  more  of the
assumptions listed in \S~\ref{sec:degen} are wrong.

\begin{figure*}
\psfig{figure=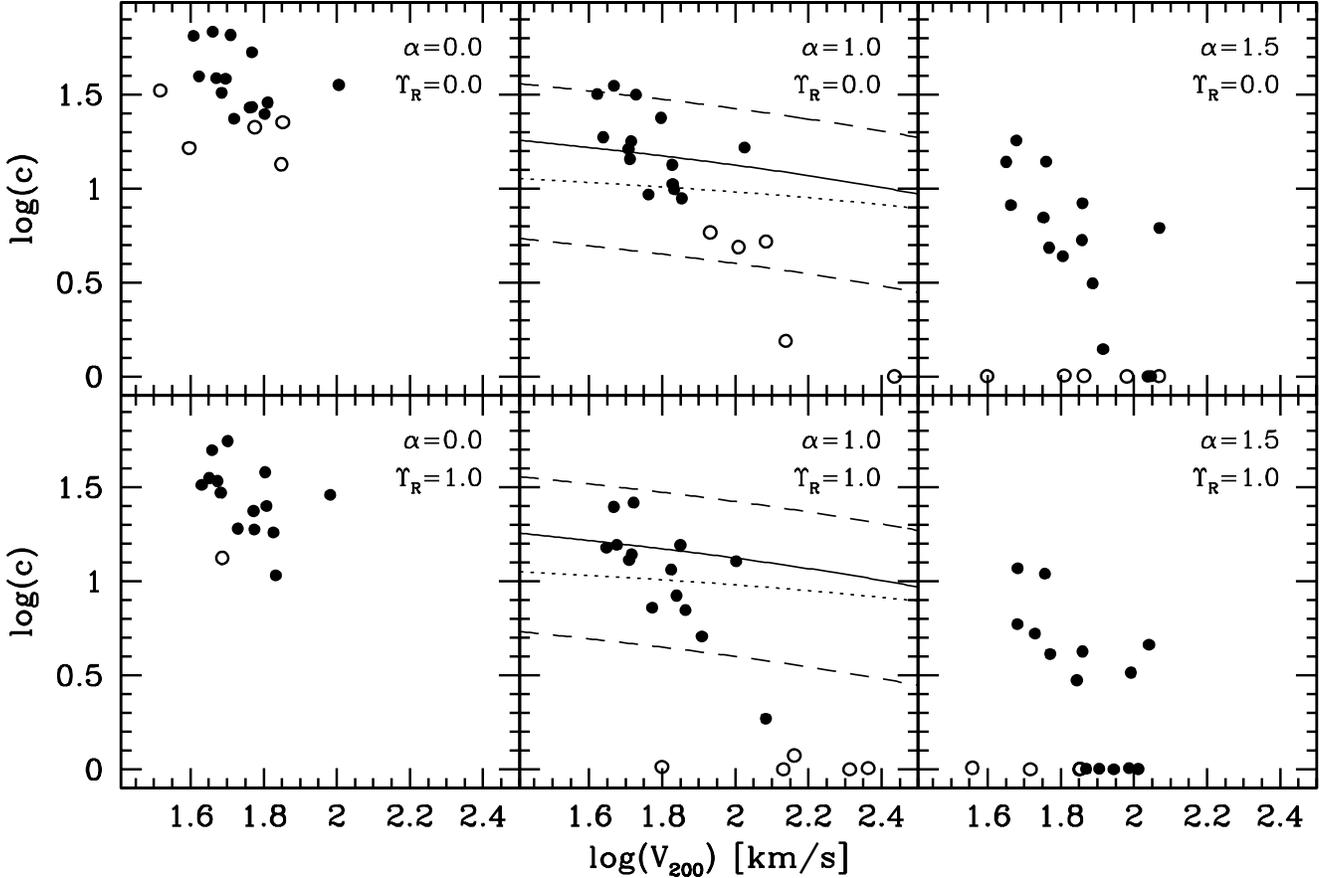,width=\hdsize}
\caption{The logarithm  of  the best-fit  concentration  parameter  as
  function  of  the logarithm  of $V_{200}$   of  the best-fit  model.
  Results are shown for two different stellar mass-to-light ratios and
  three  different values  of $\alpha$  (as indicated  in the panels).
  Solid  circles correspond to  galaxies that  are consistent with the
  $\Lambda$CDM model, whereas open circles indicate galaxies for which
  no meaningful fit  can  be obtained (indicated   by `+' and  `?'  in
  Table~\ref{tab:data}, respectively).  The solid  and dashed lines in
  the middle panels indicate the mean and the $2 \sigma$ limits of the
  distribution of halo concentrations   as predicted by the  B99 model
  for the $\Lambda$CDM cosmology.  All galaxies for which the best-fit
  models   are physically realistic  are  consistent  with this model.
  However, for $\alpha  = 1.5$ (right panels)  it  is apparent that  a
  significant fraction of the  best-fit models are unrealistic in that
  they have $c=1$).  Henceforth,  if  future simulations confirm  that
  CDM yields haloes with $\alpha \simeq 1.5$ rather than $\alpha \simeq
  1.0$,  the rotation curves analyzed  here  may signal a true problem
  for the CDM paradigm.}
\label{fig:maxc}
\end{figure*}

The  good agreement  between CDM predictions  and the  majority of the
rotation  curves   analyzed    here    is    also    illustrated    in
Figure~\ref{fig:maxc}  where we plot $c$ as  function of $V_{200}$ for
the  best-fit models with $\alpha  = 0$ (left  panels), $\alpha = 1.0$
(middle panels; see  also Table~\ref{tab:results}), and $\alpha = 1.5$
(right panels).  Results are plotted   for both $\Upsilon_R=0$  (upper
panels) and $\Upsilon_R = 1.0   \mtol$ (lower panels).  Solid  circles
correspond  to galaxies that   are consistent with  $\Lambda$CDM, open
circles to galaxies for which no meaningful fit can  be obtained.  The
solid and dashed lines in the middle  panels indicate the mean and the
$2 \sigma$  intervals of the  predictions based on  the B99 model. For
comparison, we also  plot (dotted lines) the  predictions based on the
model of Navarro, Frenk   \& White (1997)\footnote{Computed using  the
procedure outlined in their Appendix with  $f=0.01$ and $C=3.41 \times
10^3$}.  For  $\alpha  = 0.0$ and $\alpha   = 1.5$  no  such lines are
plotted, since no model predictions exist for these cases.  As already
noted  by Eke, Navarro  \&  Steinmetz (2000),  the  B99 model predicts
concentration values a factor $\sim 1.6$  larger at $V_{200}=30 \kms$,
and a slightly steeper mass dependence. Since the disagreement between
the two models is relatively small compared to the expected scatter in
$c$, our conclusions  are not sensitive to the  fact that we focus our
discussion on the B99 model. It is  apparent that for all galaxies for
which a meaningful  fit is obtained, the  best fit values for  $c$ and
$V_{200}$ are consistent with   the expected values in a  $\Lambda$CDM
cosmology.

In total we  thus find that 14 out  of 20 galaxies are consistent with
CDM.   For the remaining 6 galaxies  no meaningful fit to the observed
rotation curves can be obtained with our mass models  for any value of
$\alpha$. Henceforth, these galaxies  are neither consistent with CDM,
nor with  any  other viable alternative.   We  thus  conclude that  at
present there is  no convincing evidence that  dwarf galaxies  (or low
surface brightness galaxies, see BRDB) have dark matter haloes that are
inconsistent with CDM.    However, we  wish  to point  out  that  this
conclusion is  based  on the presumption that  CDM  haloes have $\alpha
\simeq  1.0$.  If future simulations confirm  the results by Fukushige
\& Makino (1997), Moore \etal (1998) and Klypin \etal (2000), that CDM
produces more steeply cusped   dark matter haloes with  $\alpha  \simeq
1.5$, we would have to conclude that the rotation curves analyzed here
are  only  marginally consistent with  CDM.  This  is evident from the
plots            in    Figures~\ref{fig:slopea}-\ref{fig:slopes}   and
Figure~\ref{fig:maxc}, which show that  for $\alpha \gta 1.5$ the fits
in general become rather poor, and often unrealistic.  Furthermore, we
have  made the assumption,  based on  stellar population models,  that
$\Upsilon_R   \lta 1.1 \mtol$.    If,  however, the true mass-to-light
ratio of  the  stellar disks   is  significantly higher  and/or  dwarf
galaxies have large amounts of (centrally concentrated) molecular gas,
some of  the galaxies analyzed here will  have dark matter  haloes that
are inconsistent with CDM. 

Although we have shown that the majority of the rotation curves in our
sample  are consistent with   CDM, it  does  not  imply that  they are
inconsistent  with any  of CDM's alternatives,  such as  WDM, SFDM  or
SIDM.  In  fact,  in some cases models   with a constant  density core
provide   better  fits to the   rotation curves  than for $\alpha=1.0$
(e.g., UGC~7524 and UGC~9211).  Ultimately, one might  hope to use the
rotation curves of  (dwarf) galaxies  to  put some  constraints on the
central phase space   densities or  core radii  of  their dark  matter
haloes.  This in turn  constrains the masses and/or  interaction cross
sections of the dark  matter  particles.  Unfortunately, the  rotation
curves analyzed  here do not  put  any significant constraints  on the
actual nature of  the dark matter: they are  consistent with  any dark
matter species that yields haloes with  $0 \lta \alpha  \lta 1.5$.  In
order for the rotation  curves  to put  stringent constraints  on  the
nature of the dark matter, we have to be able to much better constrain
the density distribution of  the dark  matter haloes.  However,  given
the numerous  potential sources  of  systematic  errors and both   the
mass-to-light ratio degeneracy  and the cusp-core degeneracy discussed
above it seems unlikely that HI rotation curves alone  will be able to
provide any significant constraints on the nature of the dark matter.

\subsection{Comparison with previous work}
\label{sec:comp}

Our  conclusion    that the rotation  curves  of   dwarf  galaxies are
consistent with a  $\Lambda$CDM  cosmology is  at  odds  with previous
studies   (Moore  1994;    Navarro  \etal  1996;  Moore \etal   1999b;
Blais-Ouellette, Amram \& Carignan 2000).   There are several  reasons
that may  contribute to this discrepancy.   First of all, the previous
studies have mostly used the same, small sample of dwarf galaxies, the
rotation curves of  which have been determined  in different ways.  We
have   used high-resolution HI  rotation  curves for a new, relatively
large sample of dwarf galaxies for which the rotation curves have been
derived in a  uniform way.  Furtermore, some  of the dwarf galaxies in
the  previous studies have   high inclinations or  strong non-circular
motions e.g., as the  result of a bar.   Rotation curves  derived from
such galaxies are likely to suffer from  systematic effects.  For this
reason, we excluded galaxies with  strong bars and inclinations larger
than 80$^\circ$.  In addition, we have  improved upon previous studies
by taking beam-smearing and  adiabatic  contraction into account.   As
beam-smearing can mimic the presence of a constant density core, it is
imperative  that these effects are   properly accounted for (see e.g.,
BRDB).   Also,   we  have stressed  the  importance  of  degeneracies,
systematic errors, and   the fact that not  all  models are physically
meaningful.  Finally, some of the studies that have argued against NFW
haloes fitted  their models in  some  {\it ad hoc}   manner to the last
measured data point   or the maximum of  the  rotation curve. However,
there is no reason for treating a certain data point in a special way.
In our analysis  we fitted the models to  all data points, weighted by
their errors.

\section{Summary}
\label{sec:conc }

We have analyzed high resolution HI rotation curves for a sample of 20
late-type  dwarf   galaxies.    Taking   beam-smearing  and  adiabatic
contraction into  account  we have  investigated to  what extent these
rotation curves put constraints on the (central) density distributions
of dark matter haloes.

We  have   shown  that two  distinct    degeneracies hamper  a  unique
mass-decomposition.  The first    one,  which we call  the   cusp-core
degeneracy, owes to  the fact  that  the observed rotation  curves  in
general  only  sample   the circular  velocities    of  the  system at
intermediate radii. No data is available at  either very small or very
large  radii, making it   virtually impossible to discriminate between
haloes with  a constant density core  and a $r^{-1}$  cusp.  The other
degeneracy, the mass-to-light ratio degeneracy, is well known from the
rotation curves of  (high surface  brightness)  spiral galaxies: as  a
result  of the  uncertainty  in the stellar  mass-to-light  ratio, the
relative amount of mass in the stellar  disk can be exchanged with the
amount of dark matter.   It is noteworthy  that several studies in the
past  have suggested   that dwarf galaxies   are not  impeded  by this
mass-to-light ratio degeneracy, and  are  therefore ideally suited  to
infer constraints on dark matter haloes.  However, when the effects of
beam  smearing  are taken into  account,  late-type dwarf galaxies are
also plagued by the  mass-to-light   ratio degeneracy (see  also   the
discussion in S99).

In  $\sim 70$ percent  of the  cases analyzed  here, we  find that the
rotation curves are consistent with  a $\Lambda$CDM cosmology.  In the
remaining  $\sim 30$ percent,  no   meaningful  fit  to the   observed
rotation curves could  be obtained with our mass  models for any value
of the  inner slope of the halo  density profile. Thus, although these
galaxies are inconsistent with the $\Lambda$CDM model, they can not be
considered  to  support  any  of  the  alternative dark  matter models
(unless a mechanism  can be devised that  can explain  extremely small
baryonic mass fractions). This is most likely due to systematic errors
and/or the fact that some of the assumptions underlying the models are
incorrect.  This emphasizes that care is to be taken when interpreting
rotation curve fits; sometimes   inconsistencies with CDM  predictions
are claimed without exploring the full freedom in halo parameters, and
without addressing whether or not   alternative models (i.e., with   a
constant density core) {\it can} yield realistic fits to the data.

Our main conclusion,  therefore,  is  that   there is no    convincing
evidence against  dark matter haloes  of dwarf galaxies having $r^{-1}$
cusps.  The HI rotation curves analyzed here  are consistent with dark
matter haloes with  $\alpha = 1$ and  with  concentrations as predicted
for  the currently popular $\Lambda$CDM   cosmology. Together with the
results for LSB galaxies  presented in BRDB and  S99, we thus conclude
that, based on the rotation curves of  galaxies, there is currently no
need to abandon the idea  that dark matter is  cold and collisionless. 
However,  if  future    high resolution  simulations   confirm earlier
findings of cusp slopes in the range of $\alpha = 1.5$, or if it turns
out that dwarf galaxies have disks with $\Upsilon_R \gg 1.0 \mtol$, it
may be necessary to abandon CDM in favor of an alternative that yields
haloes that are less steeply cusped. 

It is important to point out that the rotation curves studied here are
also  consistent with the presence  of dark matter haloes with constant
density  cores.     Thus,  although current  data   does  not  require
abandoning    CDM, neither does  it   allow  us  to  rule against  its
alternatives such as WDM, SFDM,  or SIDM. Discriminating between these
various dark matter models requires  rotation curves of extremely high
accuracy.  Given the numerous sources for (systematic) errors, and the
typical beam size of radio observations, we  conclude that based on HI
rotation curves  alone at best weak  limits on cosmological parameters
and/or  the nature of  the dark matter   can be obtained.  In order to
place more stringent constraints on the actual density distribution of
dark matter haloes one needs either  data of higher spatial resolution,
such as obtainable  with H$\alpha$ spectroscopy (i.e., Courteau  1997;
Blais-Ouellette,   Carignan   \&   Amram   1999;  Swaters  \etal 2000;
Blais-Ouelette \etal 2000)   or CO observations  (e.g.,
Sofue \etal 1999), and  one needs to obtain independent constraints on
the disk's  mass-to-light  ratio from for  instance  stellar  velocity
dispersions (Bottema 1993)  or, in the case  of barred disk  galaxies,
from a detailed  modelling of the  velocity fields (e.g., Englmaier \&
Gerhard 1999; Weiner, Sellwood \& Williams 2001).

%%%%%%%%%%%%%%%
% Acknowledgments
%%%%%%%%%%%%%%%

\section*{Acknowledgments}

We are grateful  to Julianne Dalcanton for  reading an earlier version
of the  paper, to  St\'ephane Charlot  for  providing results from his
stellar   population models,  and  to  the referee  for his insightful
comments.  This research was supported in part by the National Science
Foundation under  Grant No.  PHY94-07194.  FvdB  was supported by NASA
through Hubble Fellowship grant   \# HF-01102.11-97.A awarded   by the
Space Telescope Science Institute, which is  operated by AURA for NASA
under contract NAS 5-26555.

%%%%%%%%%%%%%%%
% Reference List
%%%%%%%%%%%%%%%

%%%%%%%%%%%%%%%
% Appendices
%%%%%%%%%%%%%%%

\appendix

\section[]{Comments on individual galaxies}
\label{sec:AppA}

{\bf UGC 731:} As for the models, which were moulded after this galaxy
(see  \S~\ref{sec:degen}), there  is  a large degeneracy  in the model
parameters.  The only robust results seem to be that $\alpha \lta 1.6$
and $f_{\rm bar} \lta 0.05$.  Most  importantly, the observed rotation
velocities of this galaxy are in  excellent agreement with CDM haloes,
i.e., for $\alpha = 1$, we find $c \simeq 16 - 2.3 \Upsilon_R$.

{\bf UGC 3371:}  The best-fit models prefer  a dark matter halo with a
steep   central cusp.  However,  for  a given $\Upsilon_R$ the minimum
$\chi^2$  is achieved  for $\alpha  =   \alpha_{\rm crit}$, and  these
best-fit models are therefore unrealistic.  Although $\chi^2$ seems to
depend rather strongly on $\alpha$, this owes  mainly to the extremely
small errorbars  on $V_{\rm rot}$:  the best-fit rotation  curves with
$\Upsilon_R = 1.0 \mtol$ and $\alpha = 0$, $0.5$, and $1.0$ can hardly
be discerned  by eye.  For $\alpha  = 1$ we  find $c \simeq 9.5 - 1.45
\Upsilon_R$, and we  thus  conclude that  UGC~3371 is consistent  with
CDM.

{\bf  UGC  4325:} The quality   of the fit  improves considerably with
increasing mass-to-light  ratio up  to $\Upsilon_R \simeq  7.5 \mtol$,
after which $\chi^2$ increases rapidly.  However, for $\Upsilon_R \gta
7  \mtol$  the best-fit models  have  $c=1$ and  are thus unrealistic. 
Furthermore,  S99 has shown that $0.5  \lta \Upsilon_R \lta 2.0$ based
on the stellar velocity dispersions, and $\Upsilon_R  \gta 3 \mtol$ is
unlikely  in  the light of   stellar  population models and UGC~4325's
color  of  $B-R=0.85$   (see Figure~\ref{fig:colups}).   Clearly,  the
best-fit model is not the most realistic model. As is evident from the
lower-left   panel, models   with $\Upsilon_R  =   1.0 \mtol$  provide
reasonable fits to  the data, virtually  independent  of $\alpha$. For
$\alpha = 1$ we find $c = 30.9 - 4.8 \Upsilon_R$, and we thus conclude
that UGC~4325 is consistent with CDM.

{\bf UGC 4499:} The quality of the fit depends strongly on the stellar
mass-to-light   ratio.  For  $\Upsilon_R \gta 2.0   \mtol$, the models
become unrealistic.  The best fitting  models have $\Upsilon_R = 0.0$,
which is  also unphysical.  For  $\Upsilon_R =  1 \mtol$, the best-fit
models with  $\alpha = 0.0$, $0.5$, and  $1.0$ yield virtually equally
good fits to the data, but models with  $\alpha \gta 1.1$ are excluded
by the data, since they require $c < 1$.  For $\alpha  = 1$ we find $c
\simeq  9.0   - 7.0 \Upsilon_R$.   Henceforth,  for  $\Upsilon_R = 1.0
\mtol$ UGC~4499 is inconsistent with $\Lambda$CDM  since it predicts a
too  small  $c$. Consistency  with  B99's $\Lambda$CDM  model requires
$\Upsilon_R \lta  0.7   \mtol$. Since   this is not    unrealistic, we
still consider UGC~4499 to be consistent with CDM.

{\bf   UGC   5414:}  The observed   rotation    velocities  imply that
$\Upsilon_R \lta 1.0  \mtol$. Furthermore, the  data favors a constant
density core, and is  clearly inconsistent with  CDM. However, none of
the mass-models  provides a  realistic fit, even  for  $\alpha  = 0$
(either $c < 1$ or $f_{\rm bar} < 0.01$).  

{\bf UGC 6446:} For the  models to  be realistic requires  $\Upsilon_R
\lta 2.5 \mtol$.  For $\Upsilon_R =  1.0 \mtol$ the best-fit model has
$\alpha  \simeq  1.5$.  However,  this  model  is  unrealistic  (i.e.,
$c=1$),  but models  with  $0 < \alpha   < 1.5$ all  provide virtually
equally good fits.  For $\alpha =  1.0$ we find  $c \simeq 17.4 - 10.3
\Upsilon_R$ and UGC~6446 is thus consistent with CDM.

{\bf  UGC 7232:} The properties of  this galaxy closely resemble those
of UGC~5414 and UGC~7323.  Even for $\Upsilon_R =  0$ do we not find a
best-fit model with  $c > 1$ and  $f_{\rm bar} > 0.01$.  We  therefore
conclude that no meaningful fit can be obtained for this galaxy.  Note
that the observed rotation curve consists of only five data points.

{\bf UGC 7323:} As for UGC~5414 and UGC~7232, no meaningful fit can be
obtained for this galaxy.

{\bf UGC  7399:}  In order for  the  models  to be  realistic requires
$\Upsilon_R \lta 4 \mtol$.   For $\alpha=1$ we  find $c \simeq 23.1  -
7.0 \Upsilon_R$ and UGC~7339 is thus consistent with CDM.

{\bf  UGC 7524:} This  galaxy has the best  resolved rotation curve of
all galaxies analyzed here.  Unfortunately, equation~(\ref{sbHI}) does
not yield a reasonable fit to the observed HI  surface density, and we
therefore   opted    to   use    the   full-resolution   data     (see
\S~\ref{sec:data}) to    model the   unsmeared  HI   surface   density
distribution.   The data  favors $\Upsilon_R \lta    4 \mtol$ and  low
values for $\alpha$. However, for $\alpha = 1$ we find $c \simeq 8.5 -
3.6 \Upsilon_R$, consistent with CDM.

{\bf UGC 7559:}  As is evident from the  fact that $\chi^2_{\rm red} >
18$, none of our mass-models is able to yield  a reasonable fit to the
observed rotation curve. However, this is not too surprising since for
$r  \gta 0.9$ kpc the velocity  field is highly  asymmetric (see S99),
which is not  properly reflected  by  the errorbars. As for  UGC~5414,
UGC~7232 and UGC~7323 no meaningful fit can be obtained.

{\bf UGC  7577:} No  results are   plotted for this  galaxy,  since no
model-fit was  found to  converge.   We thus classify  this  galaxy as
UGC~5414, UGC~7232, UGC~7323, and UGC~7603, in  that no meaningful fit
can be obtained. S99 suggested that  UGC~7577, which is at a projected
distance of only  37 kpc to NGC~4449, may  be a dwarf  galaxy that was
formed  by tidal interactions in    the HI streamers around   NGC~4449
(Hunter \etal  1998).  Such tidal  dwarf galaxies are expected to have
little or no dark matter (Barnes \& Hernquist 1992).

{\bf UGC 7603:} Similar to UGC~7524, no reasonable fit to the observed
HI surface density can  be obtained with equation~(\ref{sbHI}) and  we
use the full-resolution data to model the unsmeared HI surface density
distribution.  As   for the mass   models:  no meaningful fit   can be
obtained.

{\bf UGC 8490:} The rotation curve is well-resolved and prefers models
with a low mass-to-light ratio.  For $\alpha = 1$ we find $c \simeq 24
- 10 \Upsilon_R$, and this galaxy is thus consistent with CDM.

{\bf UGC   9211:}  Models   with $\Upsilon_R  =  1.0   \mtol$  provide
reasonably good  fits to the  data, virtually independent of $\alpha$. 
For $\alpha =  1.0$ we find  $c \simeq 18.3 -  3.6 \Upsilon_R$, and we
thus conclude that this galaxy is consistent with CDM.

{\bf UGC 11707:} This galaxy reveals a very large amount of freedom in
its model  parameters.  This is   partly due to the  relatively  large
errorbars for the inner data points, and which  is a reflection of the
asymmetry between the  receding and approaching rotation velocities at
$r \lta  7$ kpc (see  S99). For $\alpha  = 1.0$ one obtains  $c \simeq
13.0 - 1.8   \Upsilon_R$, and we   thus conclude that this   galaxy is
consistent with CDM.

{\bf   UGC 11861:}  This   is  one of     the few galaxies  for  which
$\chi^2_{\rm red}$  decreases with  increasing $\Upsilon_R$ (see  also
UGC~5424, UGC~8490, and, to a  lesser extent, UGC~7524 and UGC~12732). 
For $\Upsilon_R  \gta 3 \mtol$   the implied baryon  fraction  becomes
unrealistically  large.  For $\alpha  = 1$ we   find $c \simeq  16 - 3
\Upsilon_R$, and we thus conclude  that this galaxy is consistent with
CDM.

{\bf UGC 12060:} The rotation  curve of this  galaxy is well fitted by
the mass models.  As for UGC~11707, there is a large amount of freedom
in the model parameters.  For $\alpha = 1$ we obtain  $c \simeq 34.6 -
9.8 \Upsilon_R$.  This   implies  that  according  to our   definition
UGC~12060 is inconsistent  with $\Lambda$CDM for $\Upsilon_R=0$, since
the best fit halo concentration is {\it too large} (i.e.,  $c = 34.6 >
c_{\rm max} = 31.9$). However, for more realistic mass-to-light ratios
the best-fit  halo concentration is  in excellent  agreeement with the
predictions of the B99 model, as we  thus conclude that this galaxy is
consistent with CDM.

{\bf UGC 12632:}  The properties of  UGC~12632 are similar to those of
UGC~731.  For $\alpha = 1.0$ we find $c \simeq 14.0 - 1.3 \Upsilon_R$,
and we thus conclude that this galaxy is consistent with CDM.
 
{\bf UGC 12732:} For  this  galaxy, larger mass-to-light ratios  imply
smaller  values for $\alpha$.  For    $\Upsilon_R \gta 3.5 \mtol$  the
resulting  baryon    fraction  becomes  unrealistically   small.   For
$\Upsilon_R = 1.0 \mtol$, the best-fitting model has $\alpha=1.5$, but
also $c=1$, and  is thus unrealistic.  For $\alpha  = 1.0$  we find $c
\simeq 10.3 - 3.3  \Upsilon_R$, and we  thus conclude that this galaxy
is consistent with CDM.

\label{lastpage}

\clearpage

\begin{figure*}
\psfig{figure=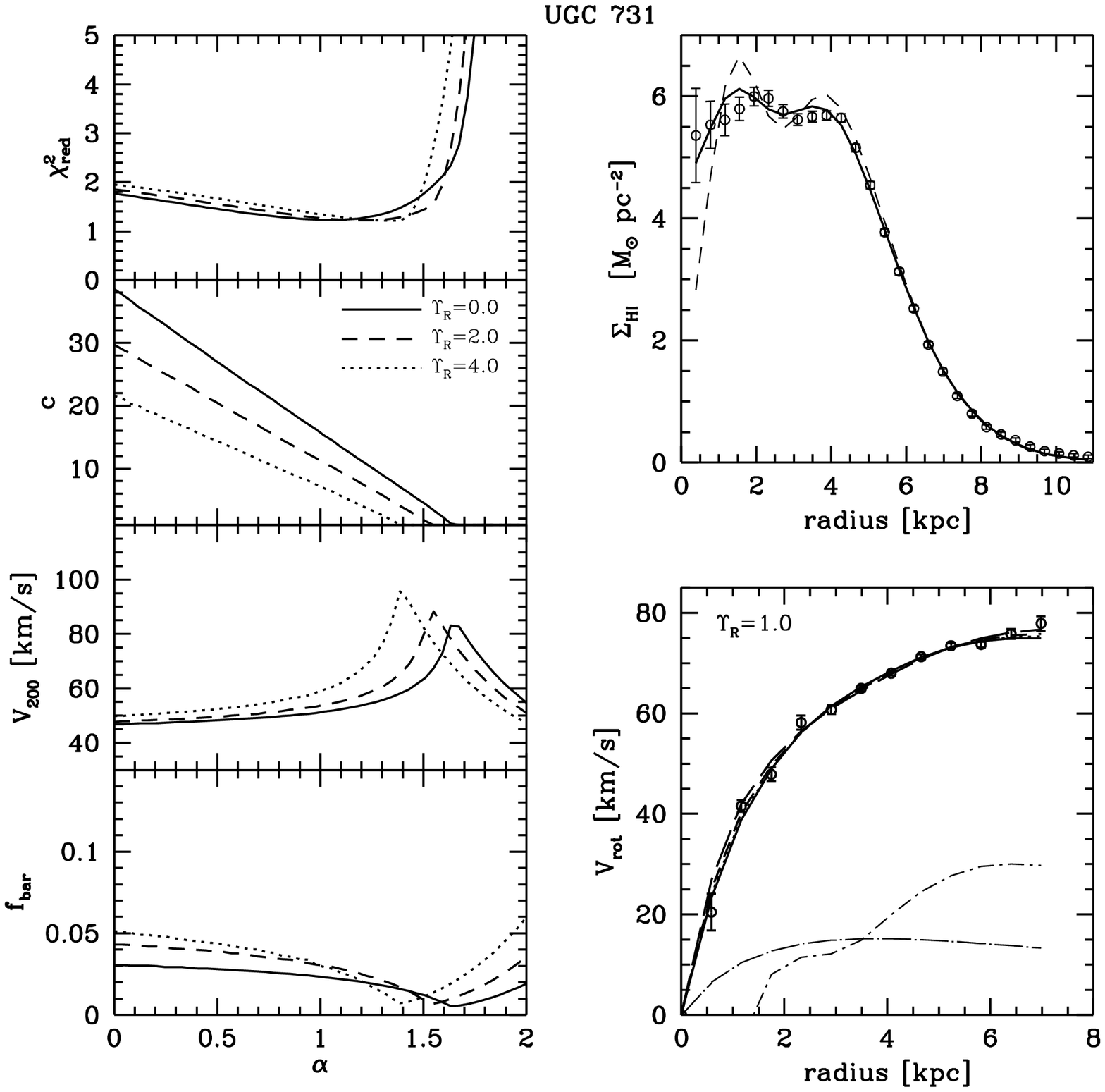,width=\hdsize}
\caption{The panel  in the upper-right   corner plots the observed  HI
  surface density (open   circles with  errorbars) together  with  the
  best-fit model of equation~(\ref{sbHI})  both before  (dashed lines)
  and after (solid lines) beam smearing. In  the cases of UGC~7524 and
  UGC~7603, were no  acceptable model fit can  be obtained, the dashed
  lines indicate  the  observed   HI   surface density at    the  full
  resolution of the observations (see \S~\ref{sec:data}), which we use
  as a model for the  unsmeared surface density  of the HI.  The  four
  panels on the left  show, from top   to bottom, $\chi^2_{\rm  red}$,
  $c$,    $V_{200}$,  and $f_{\rm bar}$  of    the best-fit models, as
  functions of  $\alpha$ and for  three different mass-to-light ratios
  (as indicated in the  second panel).  The lower-right  corner panel,
  finally,    plots the observed    rotation  curve (open circles with
  errorbars) together with four best-fit models with $\Upsilon_R = 1.0
  \mtol$: $\alpha = 0$ (solid  lines), $\alpha = 0.5$ (dotted  lines),
  $\alpha  = 1.0$ (short-dash lines)  and   $\alpha = 1.5$  (long-dash
  lines).   These four  models   only differ  in   their  dark  matter
  properties;   they  have the  same gaseous  and   stellar disks, the
  contributions of which are also indicated by (dot -- short-dash) and
  (dot -- long-dash) lines,  respectively.  These plots are useful for
  assessing the  typical quality of the model  fits  and the cusp-core
  degeneracy discussed in the text.}
\label{fig:slopea}
\end{figure*}

\clearpage

\begin{figure*}
\psfig{figure=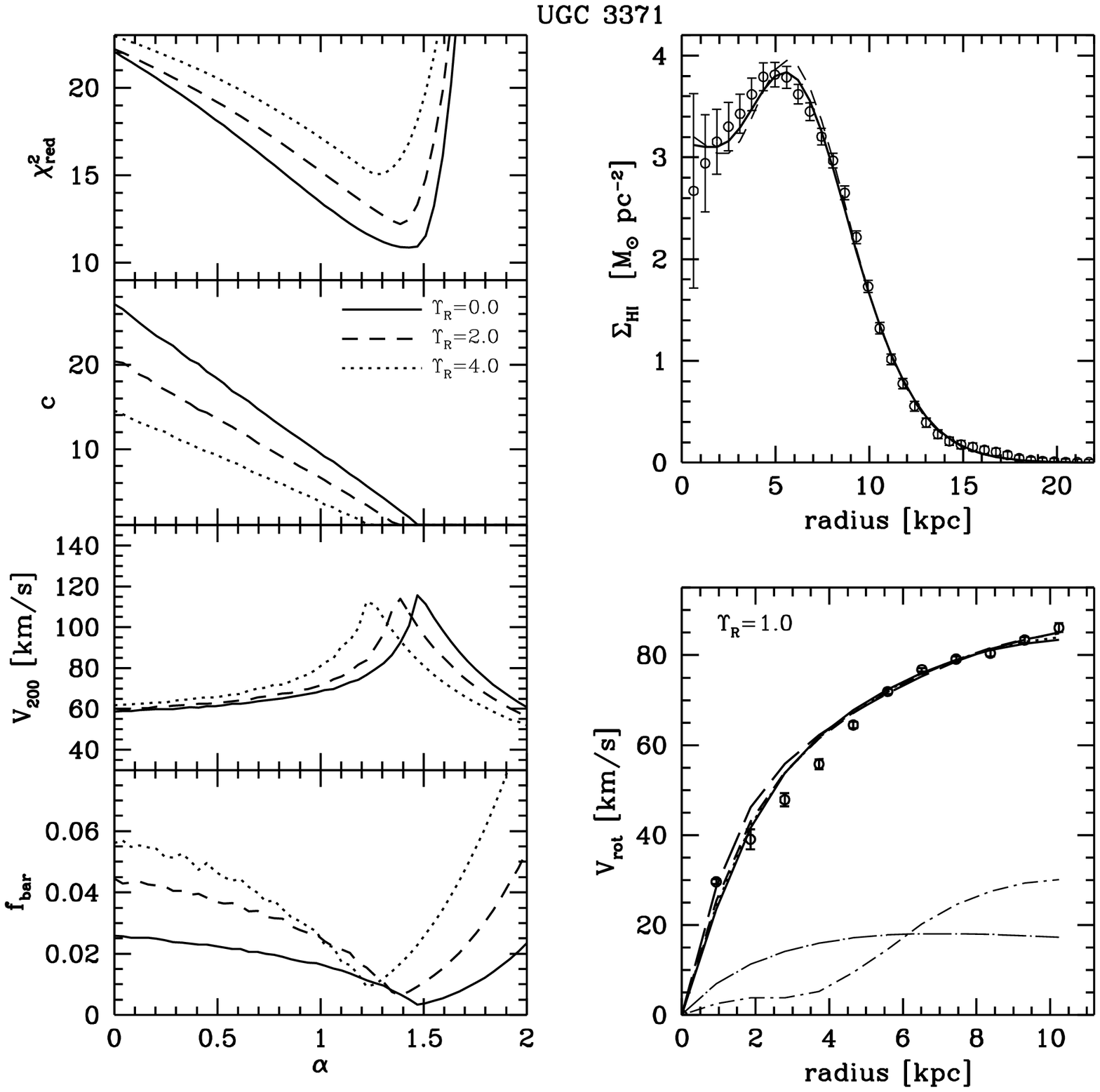,width=\hdsize}
\caption{Same as Figure~\ref{fig:slopea} but for UGC~3371}
\label{fig:slopeb}
\end{figure*}

\clearpage

\begin{figure*}
\psfig{figure=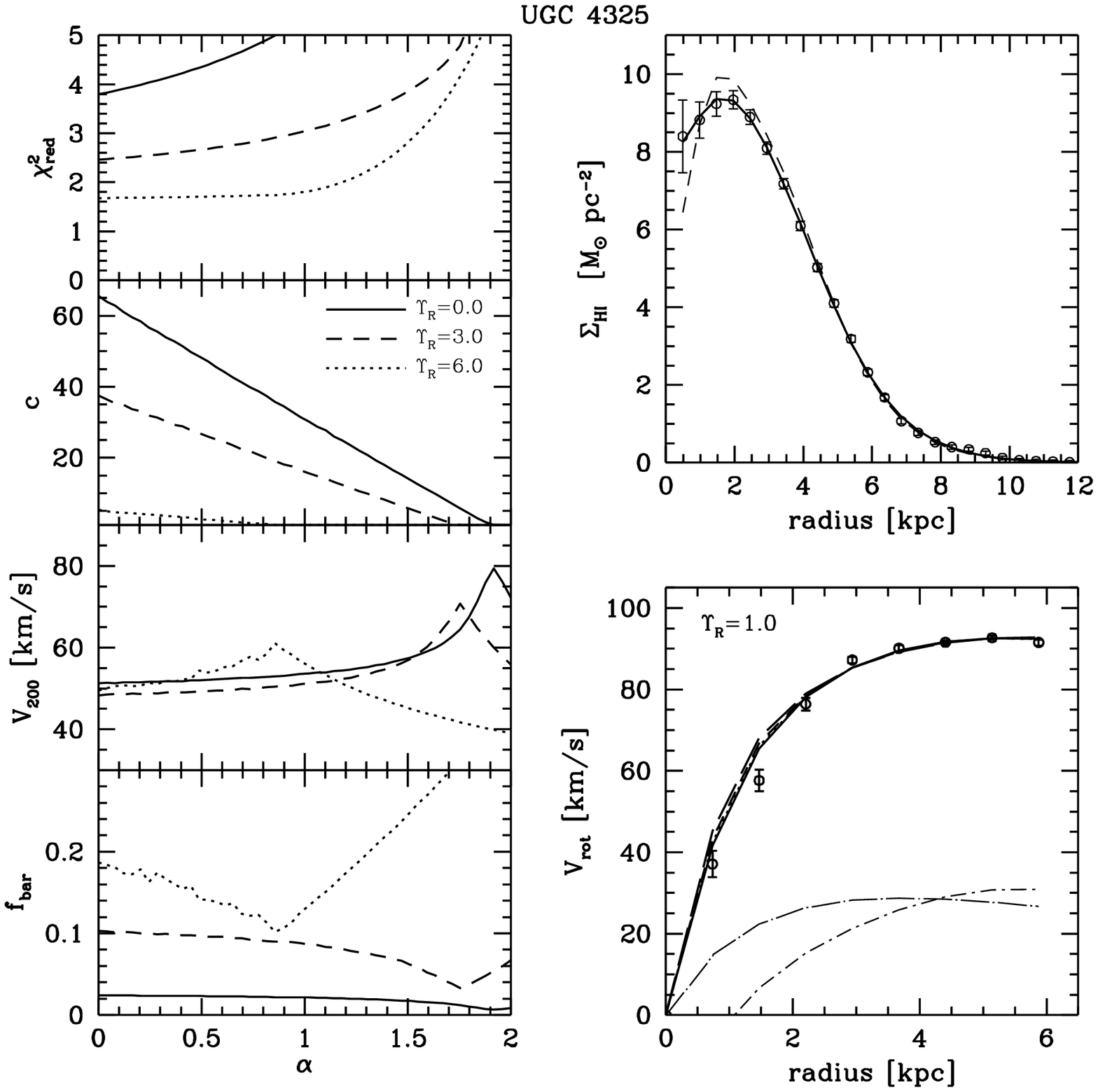,width=\hdsize}
\caption{Same as Figure~\ref{fig:slopea} but for UGC~4325}
\label{fig:slopec}
\end{figure*}

\clearpage

\begin{figure*}
\psfig{figure=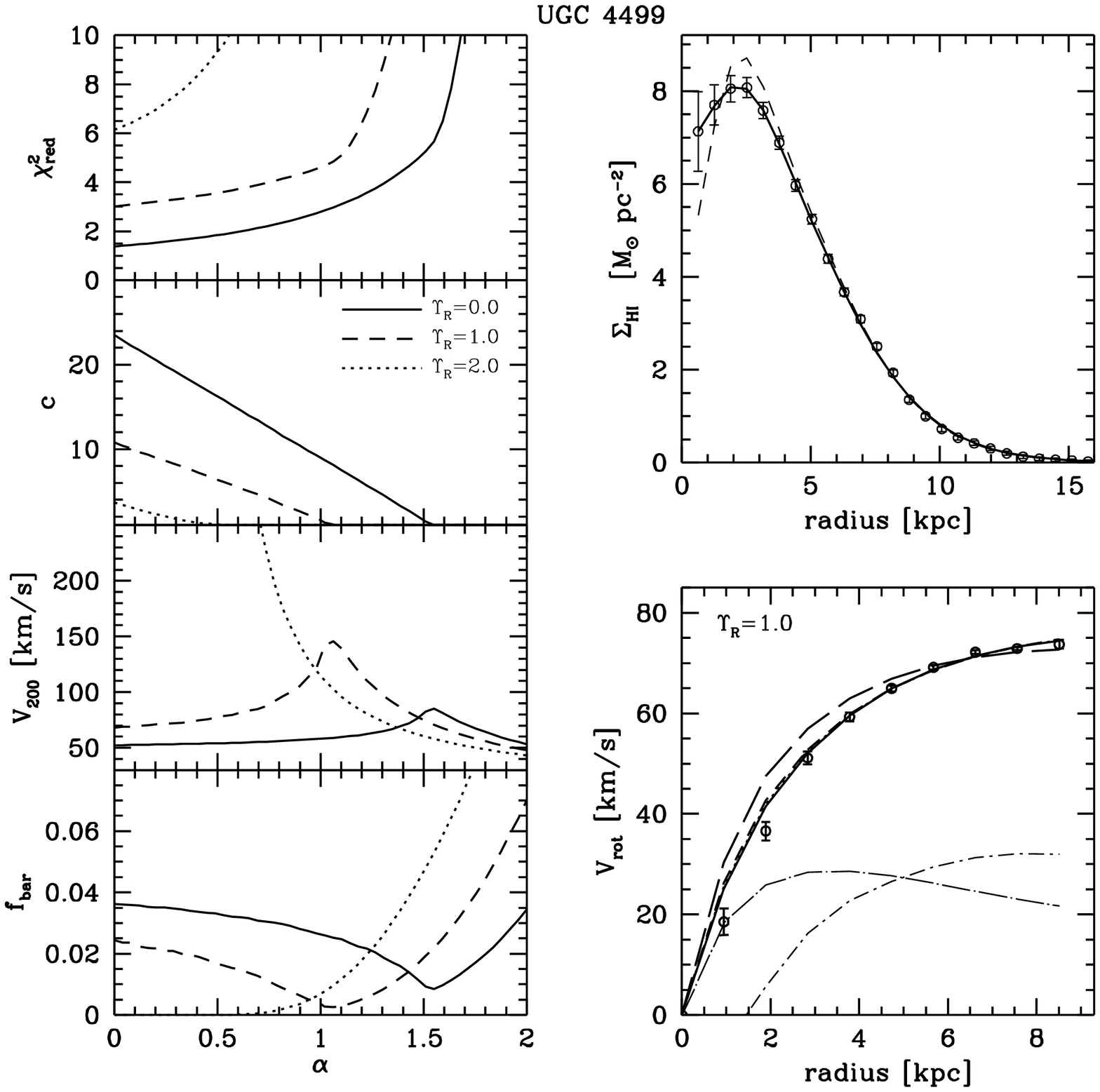,width=\hdsize}
\caption{Same as Figure~\ref{fig:slopea} but for UGC~4499}
\label{fig:sloped}
\end{figure*}

\clearpage

\begin{figure*}
\psfig{figure=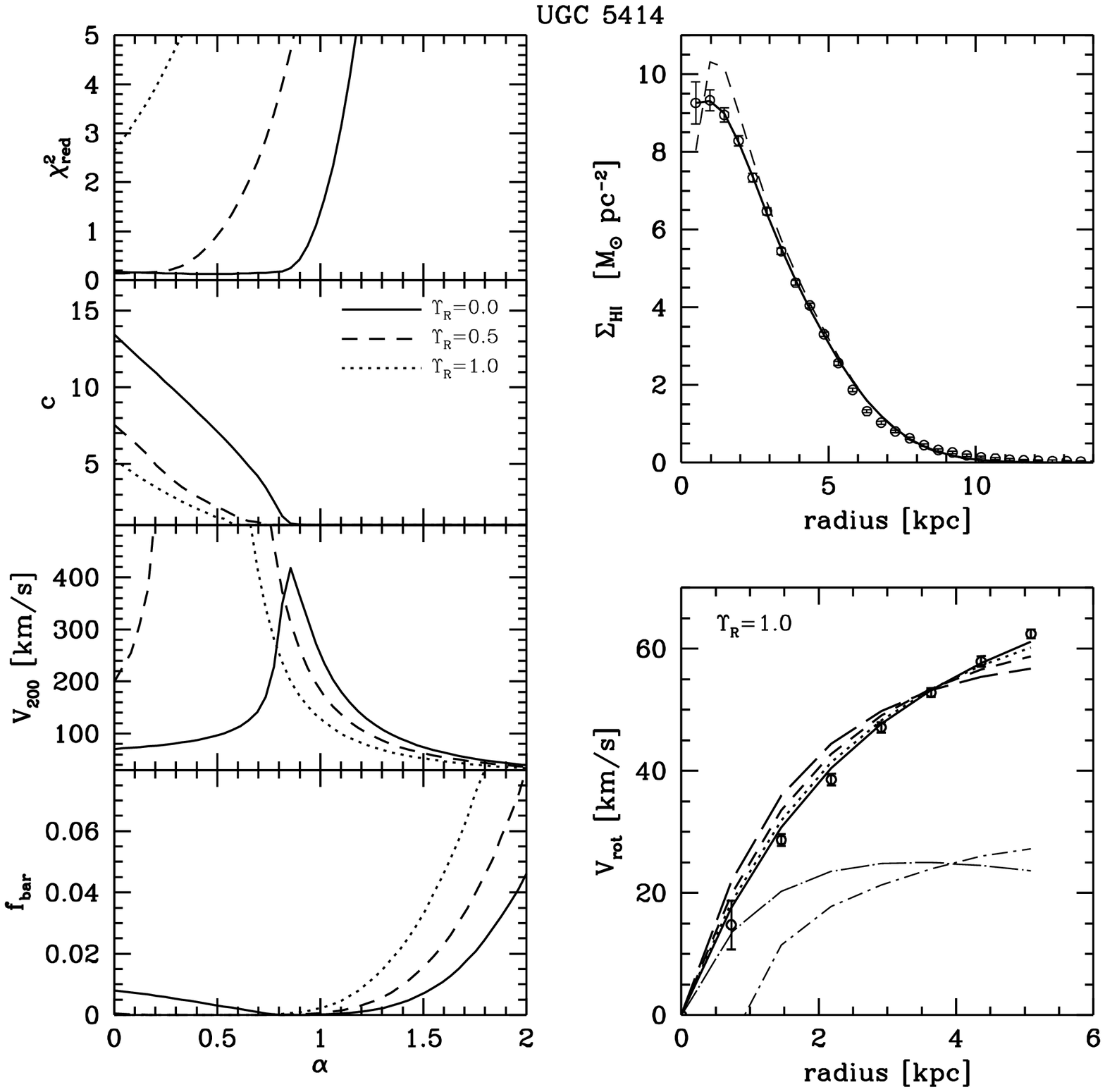,width=\hdsize}
\caption{Same as Figure~\ref{fig:slopea} but for UGC~5414}
\label{fig:slopee}
\end{figure*}

\clearpage

\begin{figure*}
\psfig{figure=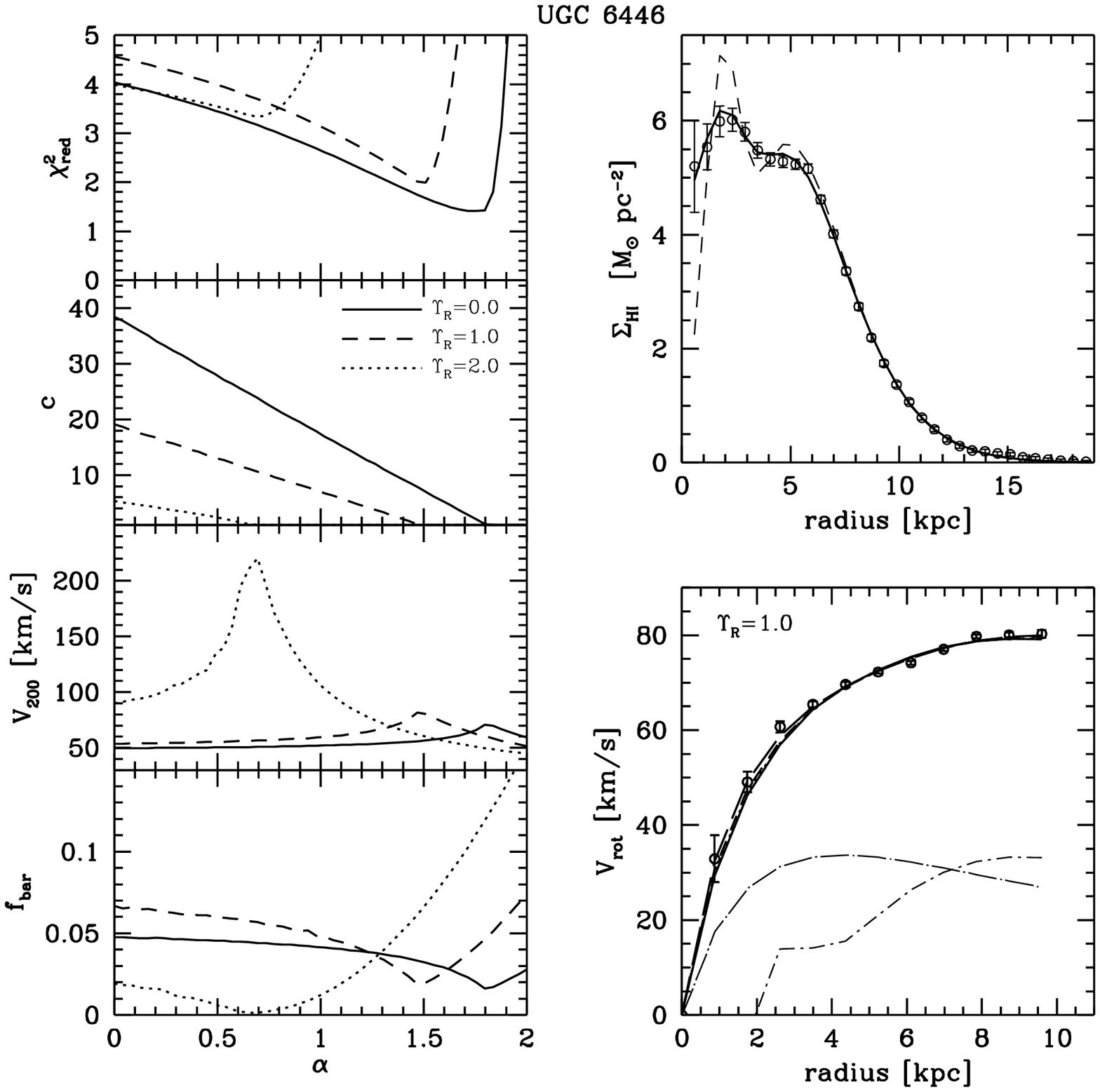,width=\hdsize}
\caption{Same as Figure~\ref{fig:slopea} but for UGC~6446}
\label{fig:slopef}
\end{figure*}

\clearpage

\begin{figure*}
\psfig{figure=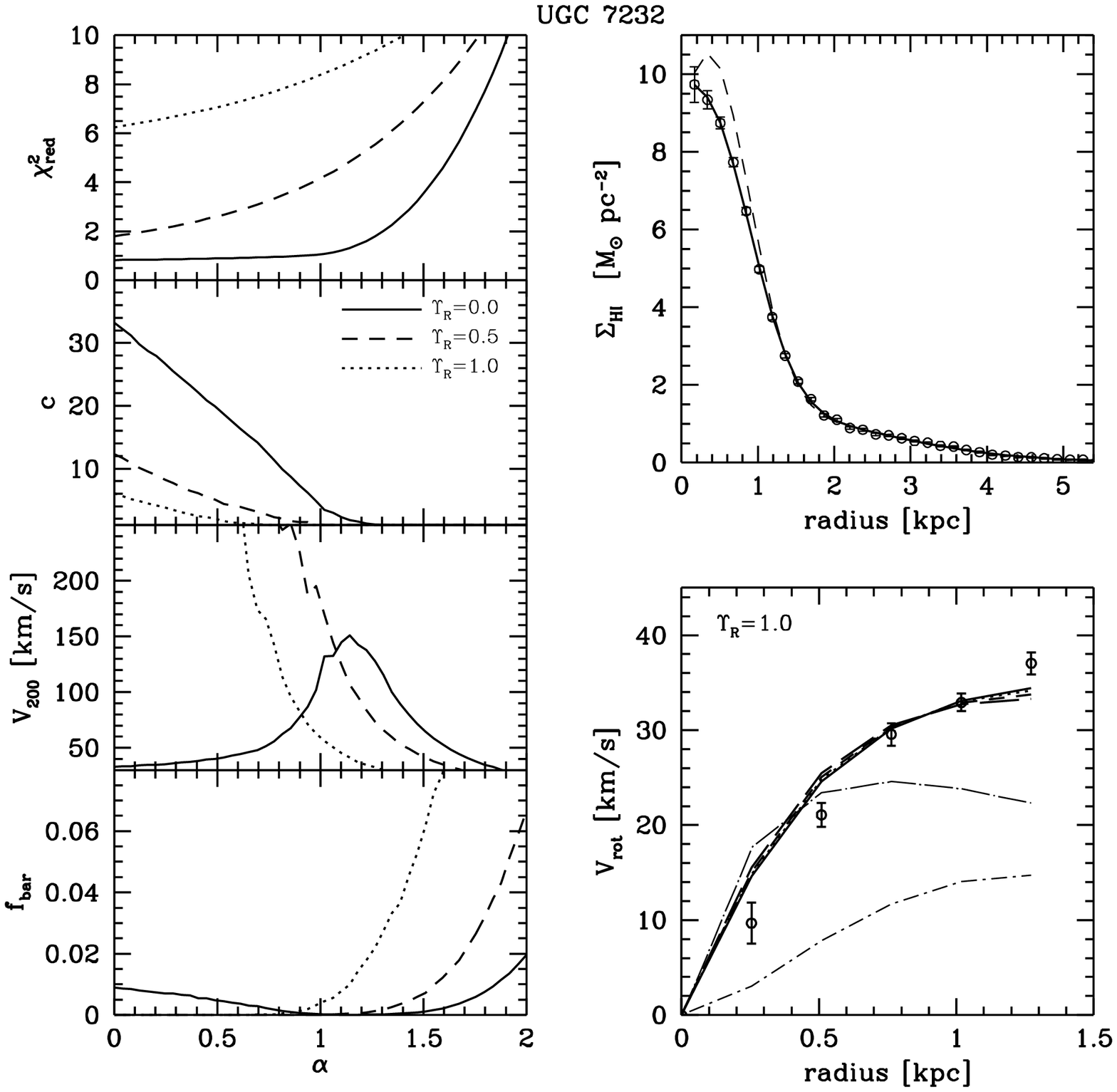,width=\hdsize}
\caption{Same as Figure~\ref{fig:slopea} but for UGC~7232}
\label{fig:slopeg}
\end{figure*}

\clearpage

\begin{figure*}
\psfig{figure=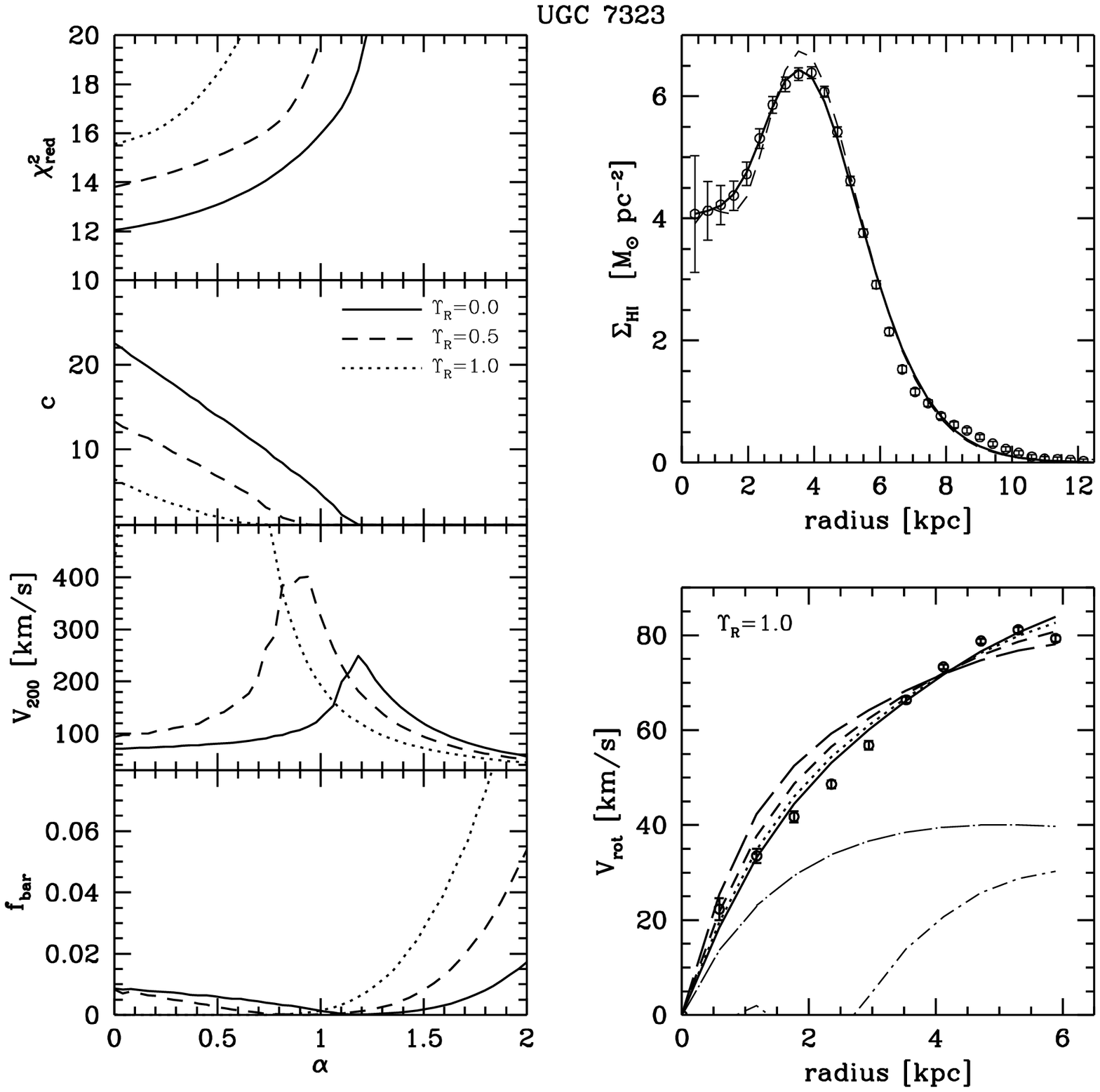,width=\hdsize}
\caption{Same as Figure~\ref{fig:slopea} but for UGC~7323}
\label{fig:slopeh}
\end{figure*}

\clearpage

\begin{figure*}
\psfig{figure=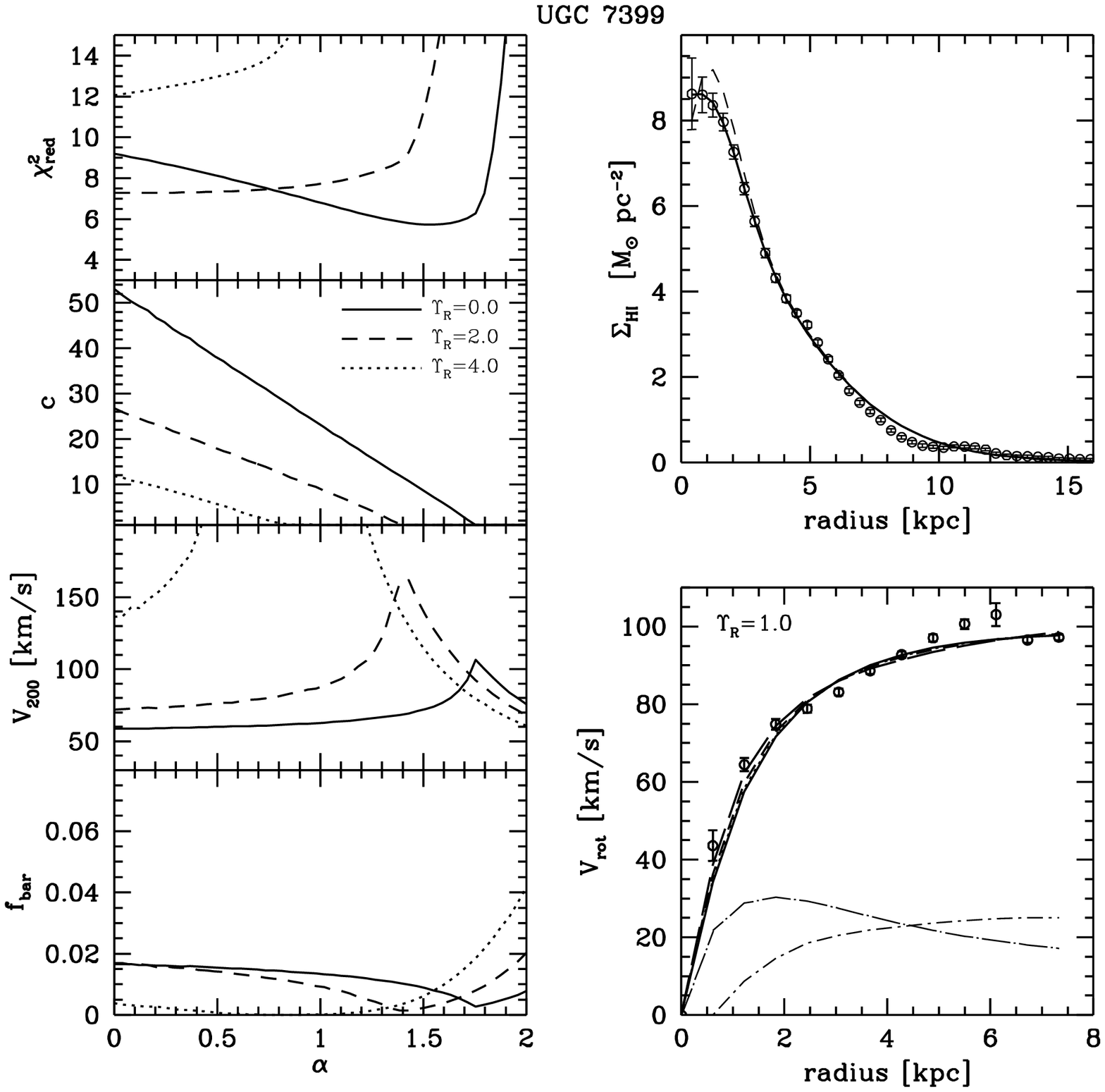,width=\hdsize}
\caption{Same as Figure~\ref{fig:slopea} but for UGC~7399}
\label{fig:slopei}
\end{figure*}

\clearpage

\begin{figure*}
\psfig{figure=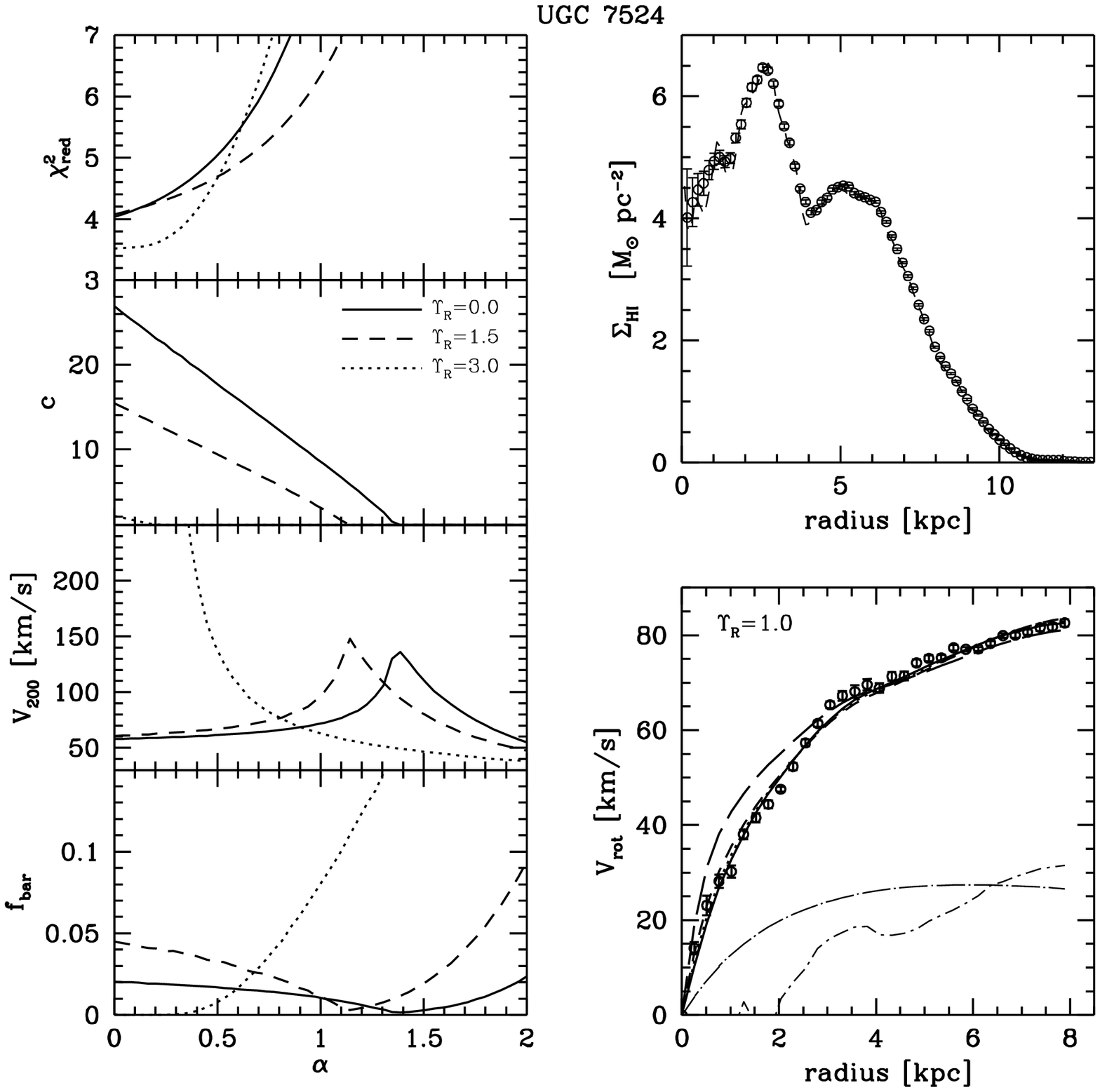,width=\hdsize}
\caption{Same as Figure~\ref{fig:slopea} but for UGC~7524}
\label{fig:slopej}
\end{figure*}

\clearpage

\begin{figure*}
\psfig{figure=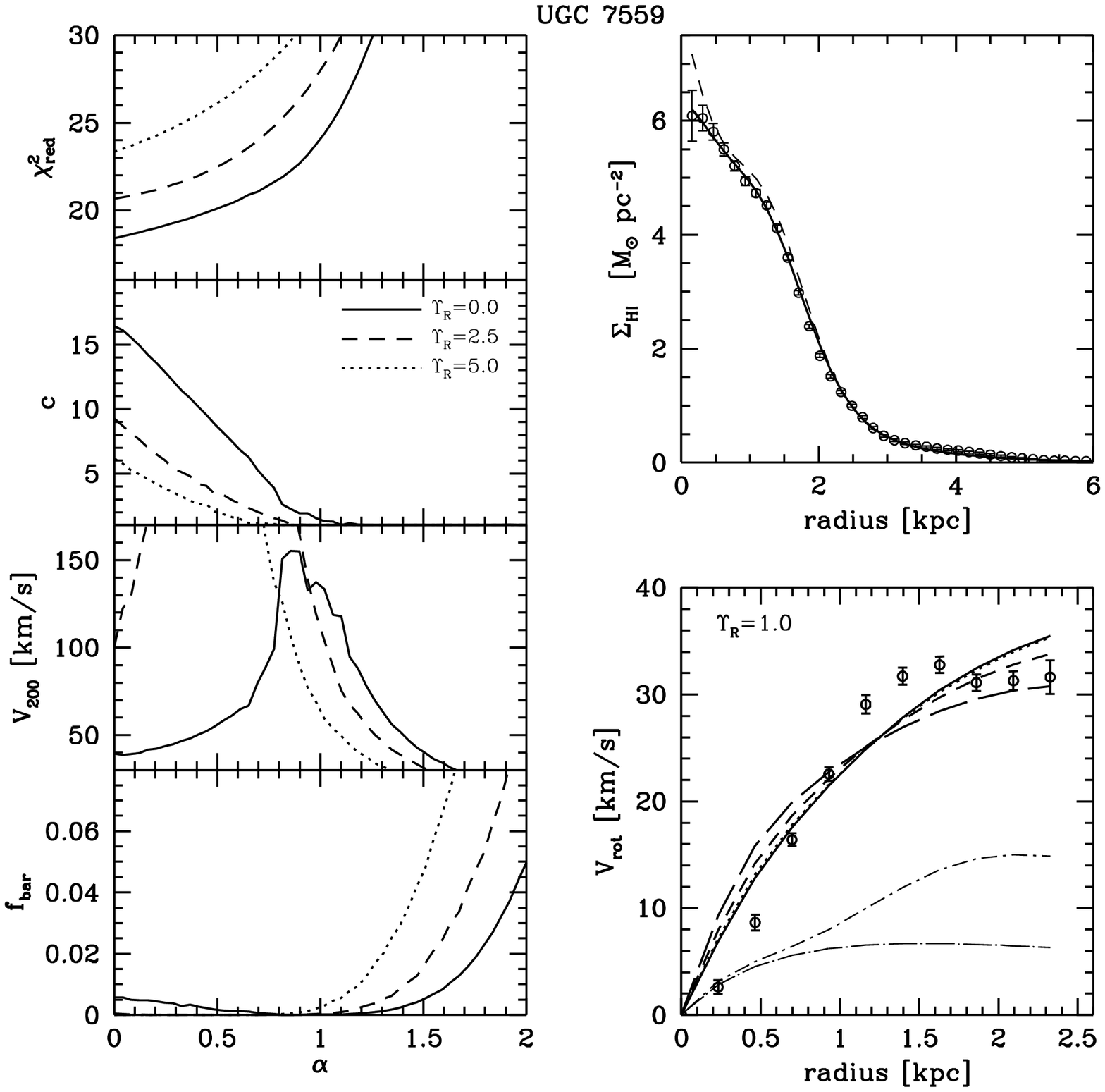,width=\hdsize}
\caption{Same as Figure~\ref{fig:slopea} but for UGC~7559}
\label{fig:slopek}
\end{figure*}

\clearpage

\begin{figure*}
\psfig{figure=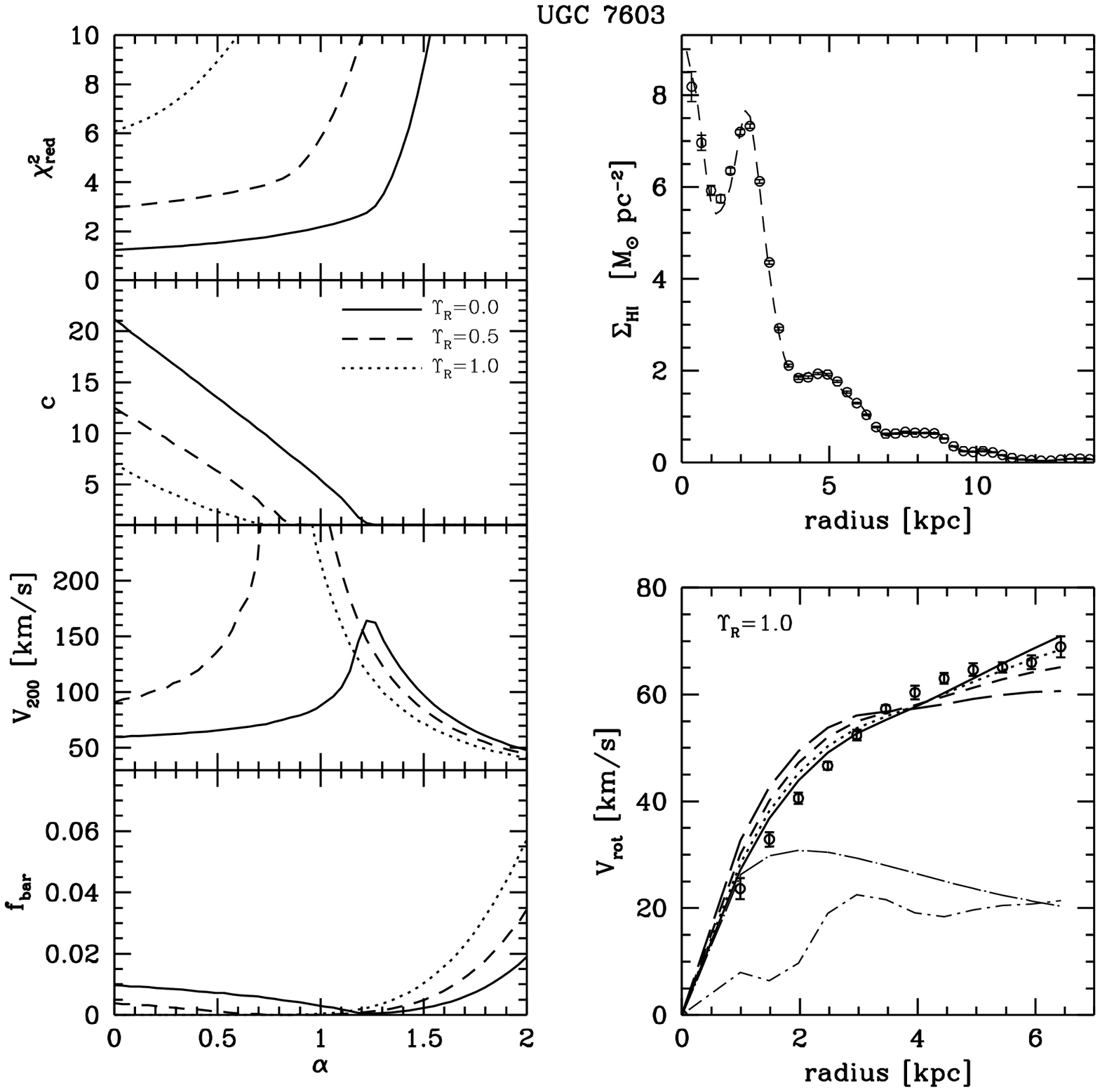,width=\hdsize}
\caption{Same as Figure~\ref{fig:slopea} but for UGC~7603}
\label{fig:slopel}
\end{figure*}

\clearpage

\begin{figure*}
\psfig{figure=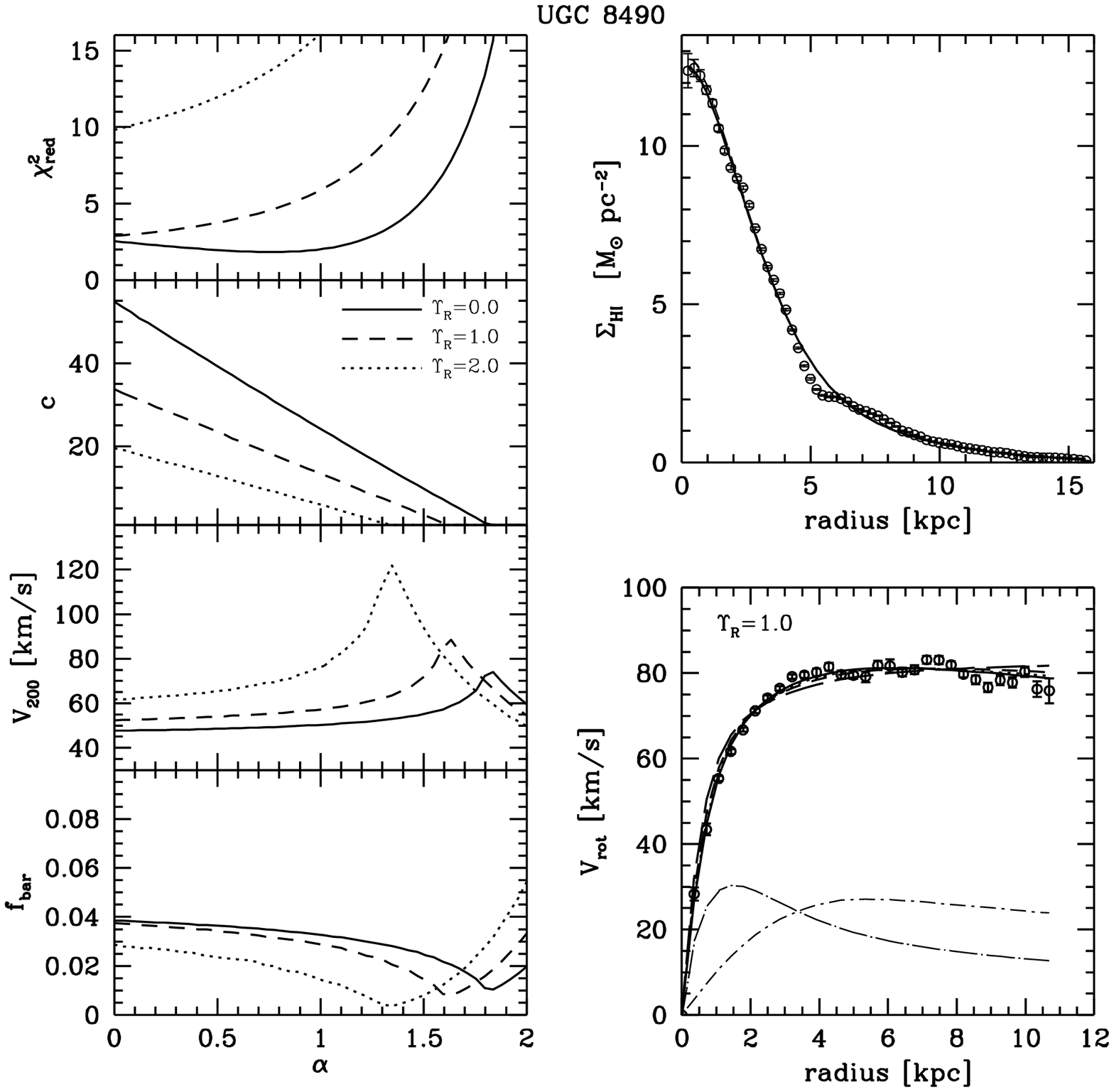,width=\hdsize}
\caption{Same as Figure~\ref{fig:slopea} but for UGC~8490}
\label{fig:slopem}
\end{figure*}

\clearpage

\begin{figure*}
\psfig{figure=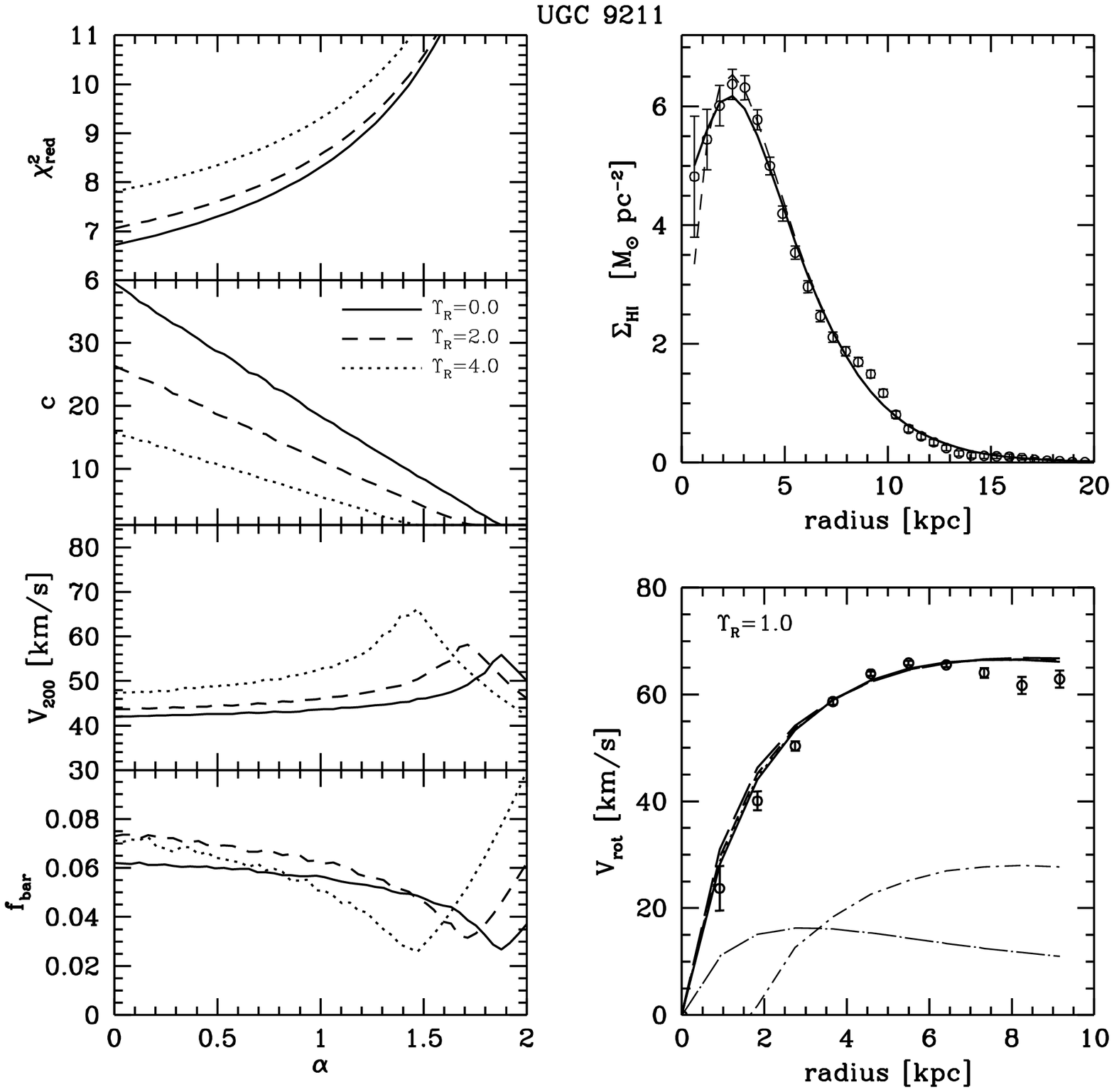,width=\hdsize}
\caption{Same as Figure~\ref{fig:slopea} but for UGC~9211}
\label{fig:slopen}
\end{figure*}

\clearpage

\begin{figure*}
\psfig{figure=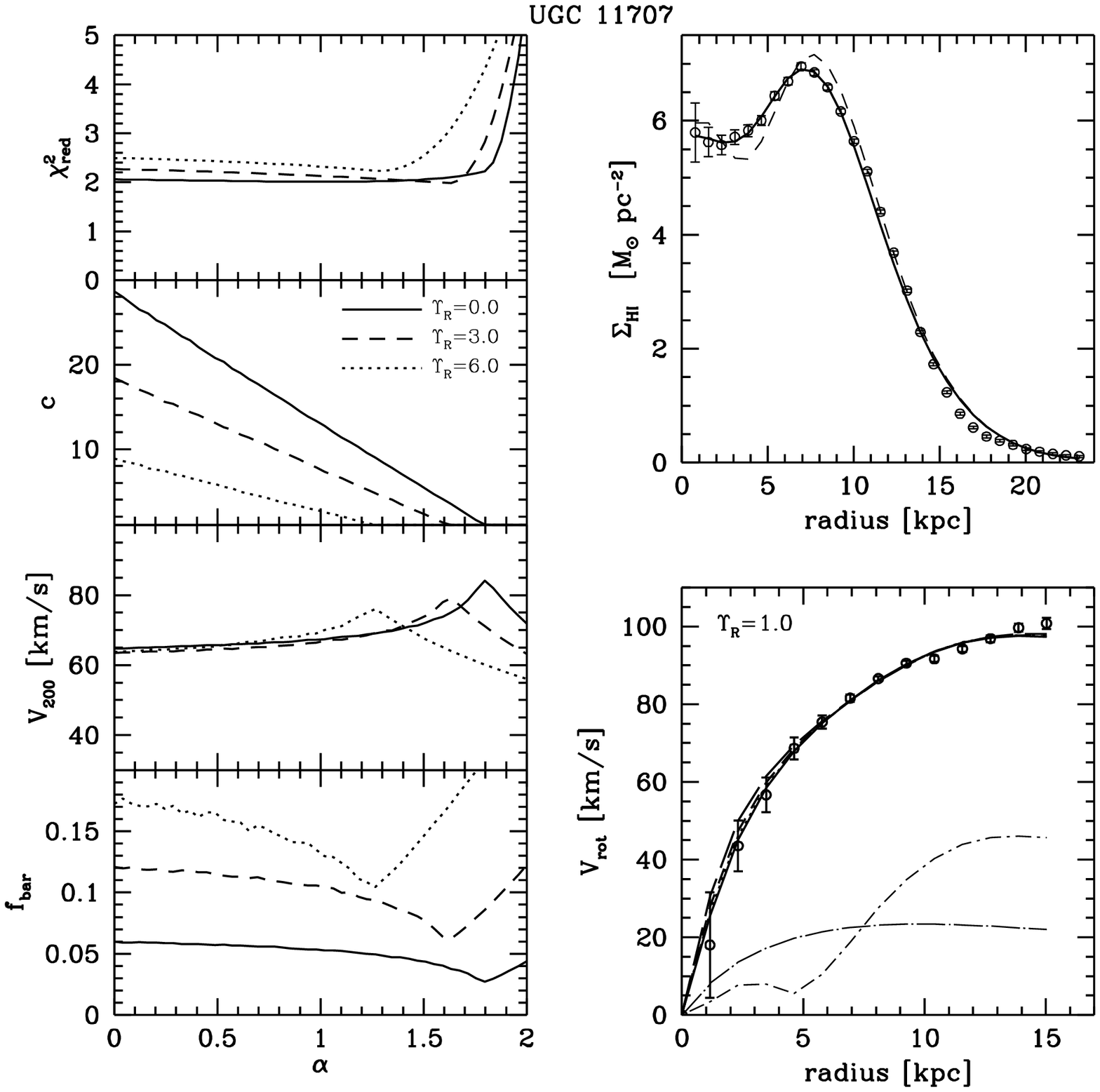,width=\hdsize}
\caption{Same as Figure~\ref{fig:slopea} but for UGC~11707}
\label{fig:slopeo}
\end{figure*}

\clearpage

\begin{figure*}
\psfig{figure=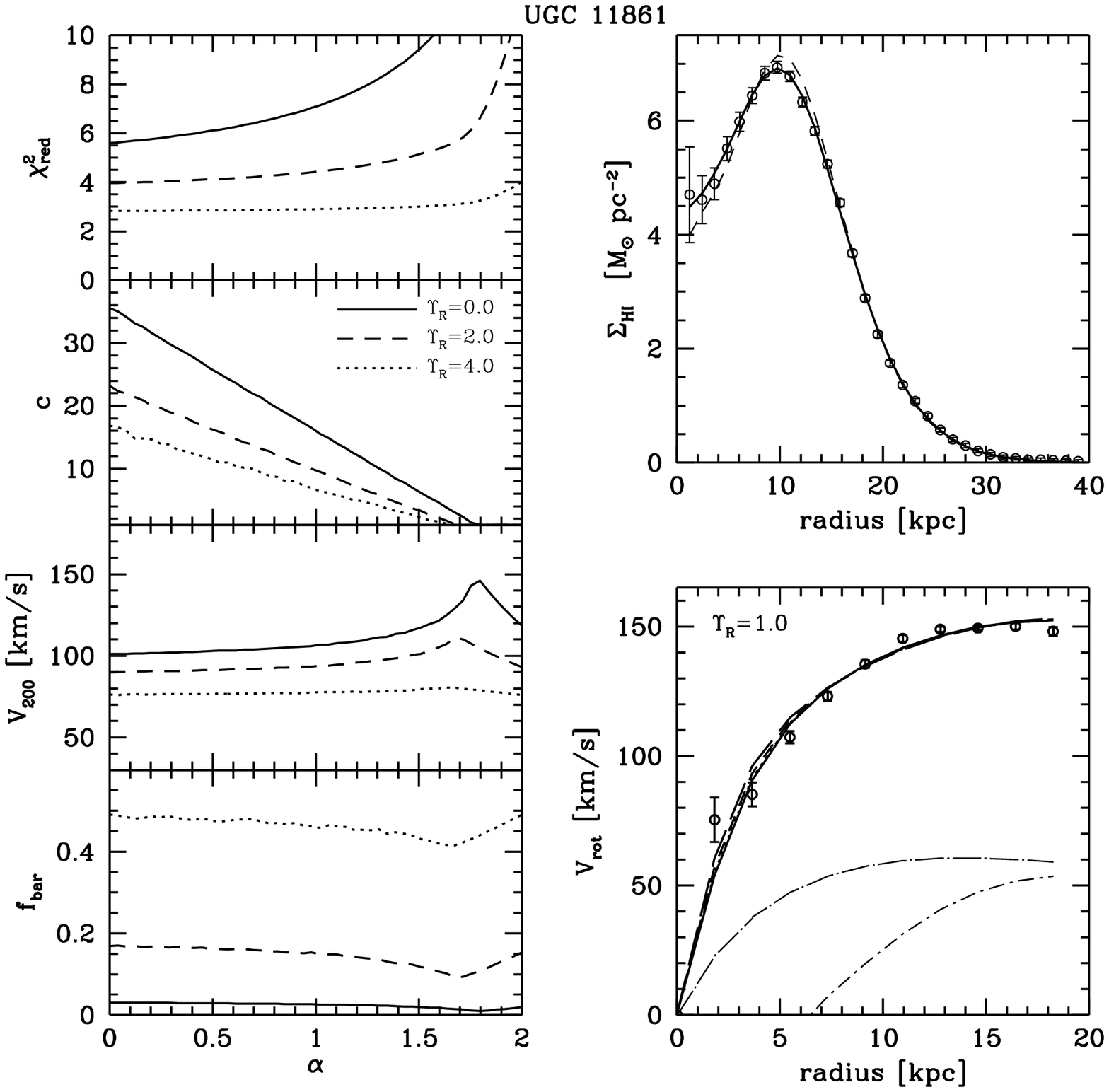,width=\hdsize}
\caption{Same as Figure~\ref{fig:slopea} but for UGC~11861}
\label{fig:slopep}
\end{figure*}

\clearpage

\begin{figure*}
\psfig{figure=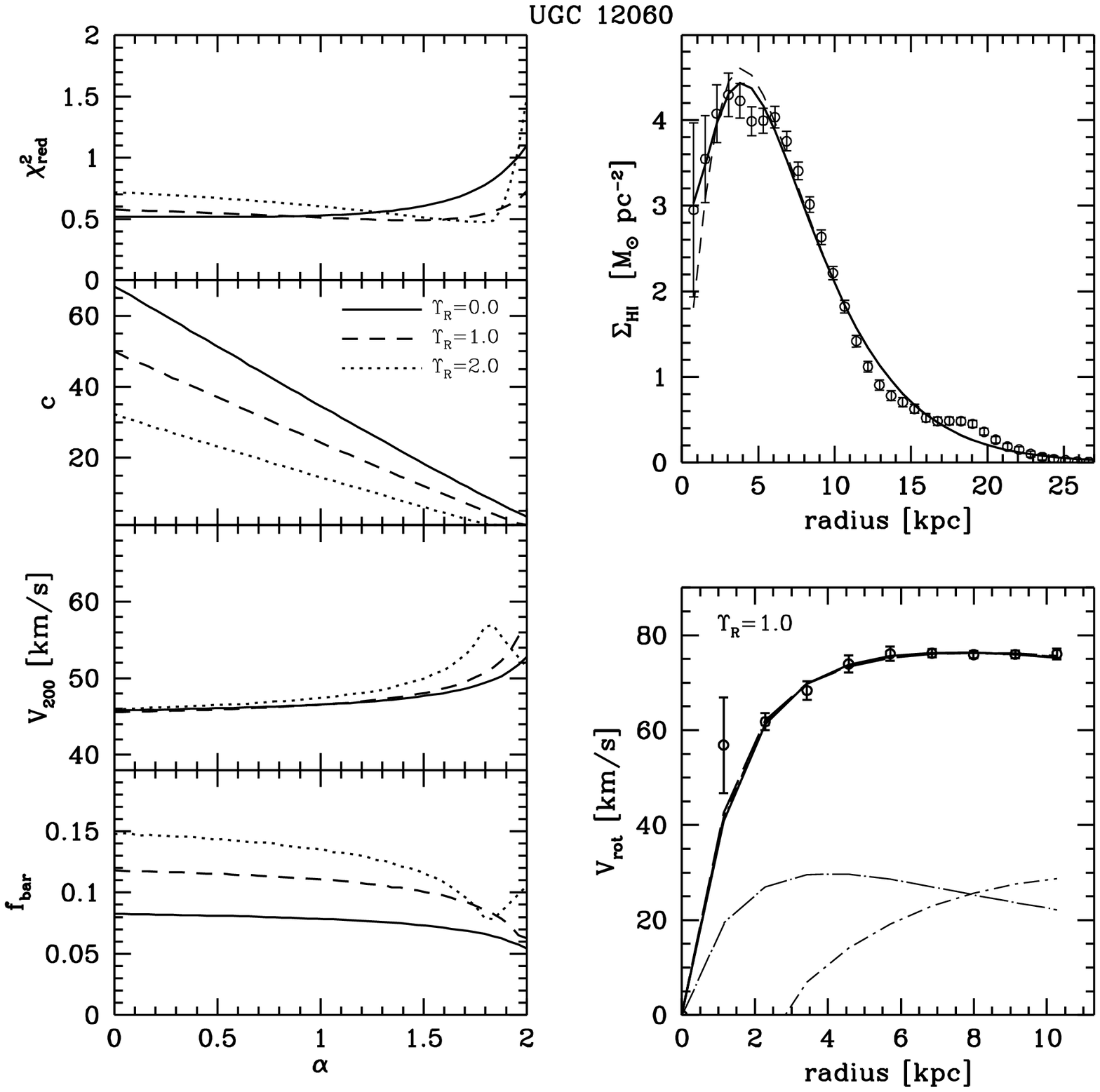,width=\hdsize}
\caption{Same as Figure~\ref{fig:slopea} but for UGC~12060}
\label{fig:slopeq}
\end{figure*}

\clearpage

\begin{figure*}
\psfig{figure=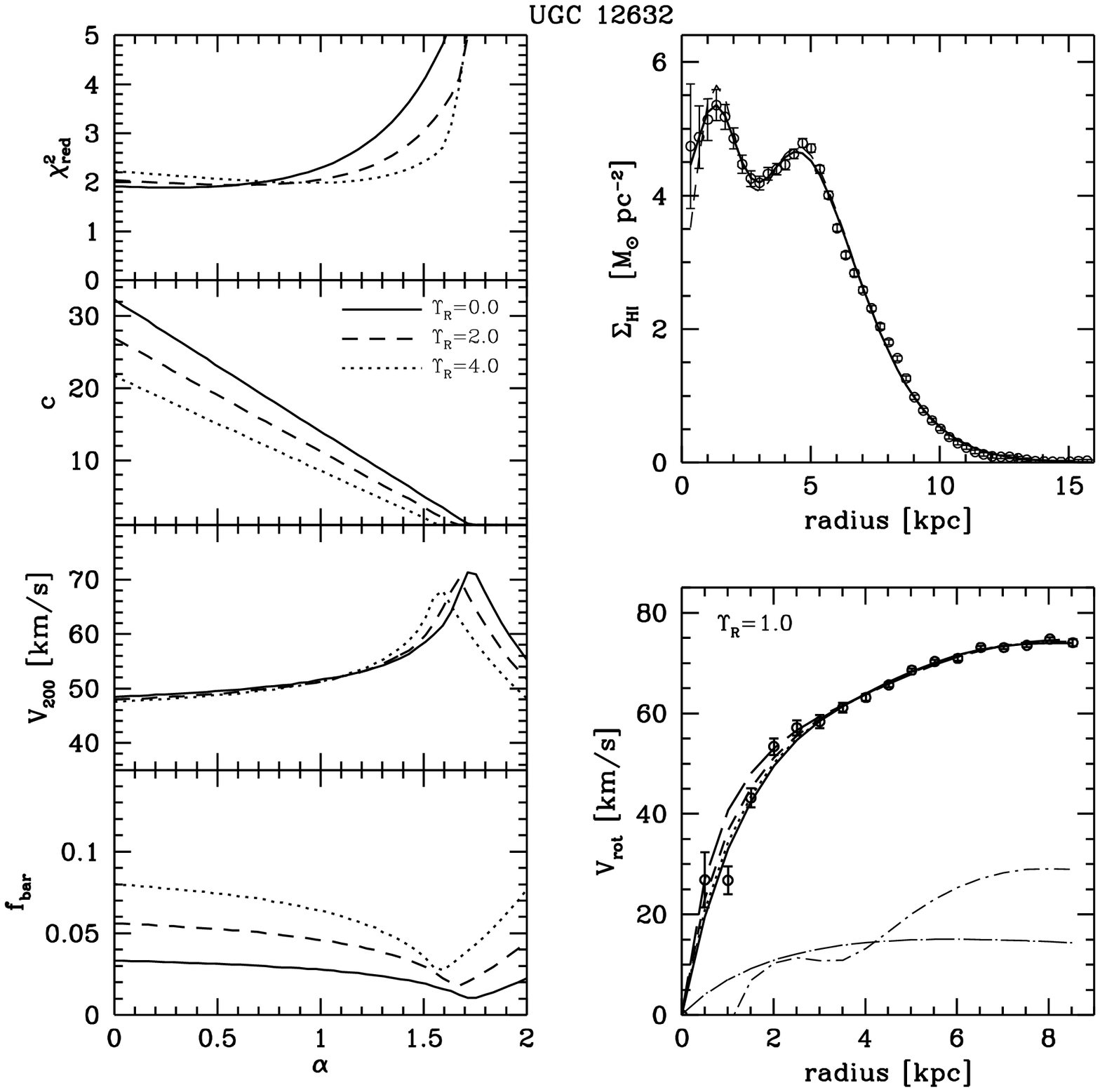,width=\hdsize}
\caption{Same as Figure~\ref{fig:slopea} but for UGC~12632}
\label{fig:sloper}
\end{figure*}

\clearpage

\begin{figure*}
\psfig{figure=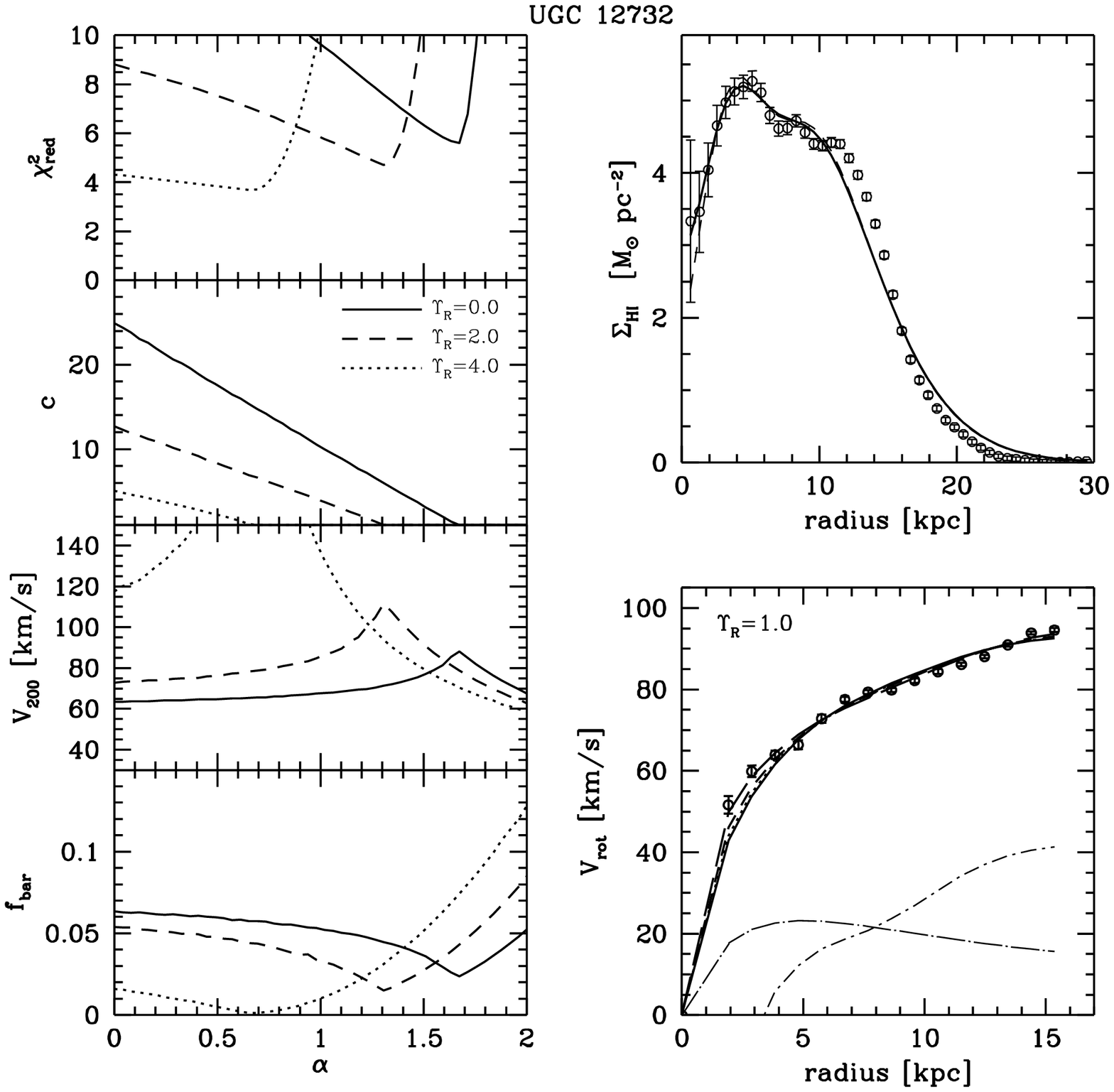,width=\hdsize}
\caption{Same as Figure~\ref{fig:slopea} but for UGC~12732}
\label{fig:slopes}
\end{figure*}

\end{document}